\documentclass[dvipsnames]{article}

\usepackage{microtype}  
\usepackage{enumitem}
\setlist[itemize]{
    leftmargin=2em,        
    rightmargin=0em,       
    itemsep=2pt,           
    parsep=0pt,            
    topsep=4pt,            
    partopsep=0pt,         
    labelsep=0.5em,        
}

\usepackage{colm2024_conference}
\usepackage{tabularx} 

\PassOptionsToPackage{table,dvipsnames}{xcolor}
\usepackage[table,dvipsnames]{xcolor}
\definecolor{lightgray}{gray}{0.9}
\definecolor{quotecolor}{RGB}{70,70,70}
\definecolor{lightpurple}{RGB}{230,230,250}

\usepackage{graphicx}
\usepackage{booktabs}
\usepackage{array}
\usepackage{tabularx}
\usepackage{colortbl}
\usepackage{multirow}
\usepackage{makecell}
\usepackage{adjustbox}
\usepackage{wrapfig}
\usepackage{caption}
\usepackage{subcaption}
\usepackage{algorithm}
\usepackage{algpseudocode}

\usepackage{amsmath, amssymb}

\usepackage{amsmath,amsfonts,bm}

\def\eqref#1{equation~\ref{#1}}

\def\1{\bm{1}}

\DeclareMathAlphabet{\mathsfit}{\encodingdefault}{\sfdefault}{m}{sl}
\SetMathAlphabet{\mathsfit}{bold}{\encodingdefault}{\sfdefault}{bx}{n}
\usepackage{url}
\usepackage{xurl}
\usepackage{nicefrac}
\usepackage{fancyvrb}
\usepackage[normalem]{ulem} 

\usepackage{CJKutf8}

\usepackage{tikz}
\usetikzlibrary{positioning}
\usepackage{pgfplots}
\pgfplotsset{compat=1.18}

\usepackage[raster,skins]{tcolorbox}
\tcbuselibrary{breakable}
\usepackage{transparent}

\usepackage{calligra}

\newtcolorbox{myquote}[1][]{
  enhanced,
  frame hidden,
  boxrule=0pt,
  arc=5pt,
  width=0.9\textwidth,
  before skip=5pt,
  after skip=10pt,
  boxsep=15pt,
  left=15pt, right=15pt, top=5pt, bottom=8pt,
  colback=lightpurple, opacityback=0.1,
  drop fuzzy shadow=lightpurple!60,
  sharp corners,
  #1
}

\title{
XekRung Technical Report}

\author{
 \vspace{15pt}
 Alibaba Security AGI Lab
 \vspace{10pt}}

\begin{document}

\begin{CJK}{UTF8}{gbsn}
\maketitle


\begin{abstract}

We present XekRung, a frontier large language model for cybersecurity, designed to provide comprehensive security capabilities. 
To achieve this, we develop diverse data synthesis pipelines tailored to the cybersecurity domain, enabling the scalable construction of high-quality training data and providing a strong foundation for cybersecurity knowledge and understanding. 
Building on this foundation, we establish a complete training pipeline spanning continued pre-training (CPT), supervised fine-tuning (SFT), and reinforcement learning (RL) to further extend the model’s capabilities. 
We further introduce a multi-dimensional evaluation system to guide the iterative improvement of both domain-specific and general-purpose abilities. 
Extensive experiments demonstrate that XekRung achieves state-of-the-art performance on cybersecurity-specific benchmarks among models of the same scale, while maintaining strong performance on general benchmarks.
\end{abstract}



\begin{figure}[!ht]
    \centering
    \includegraphics[
        width=1.0\linewidth,
        trim=0 100 0 0,
        clip
    ]{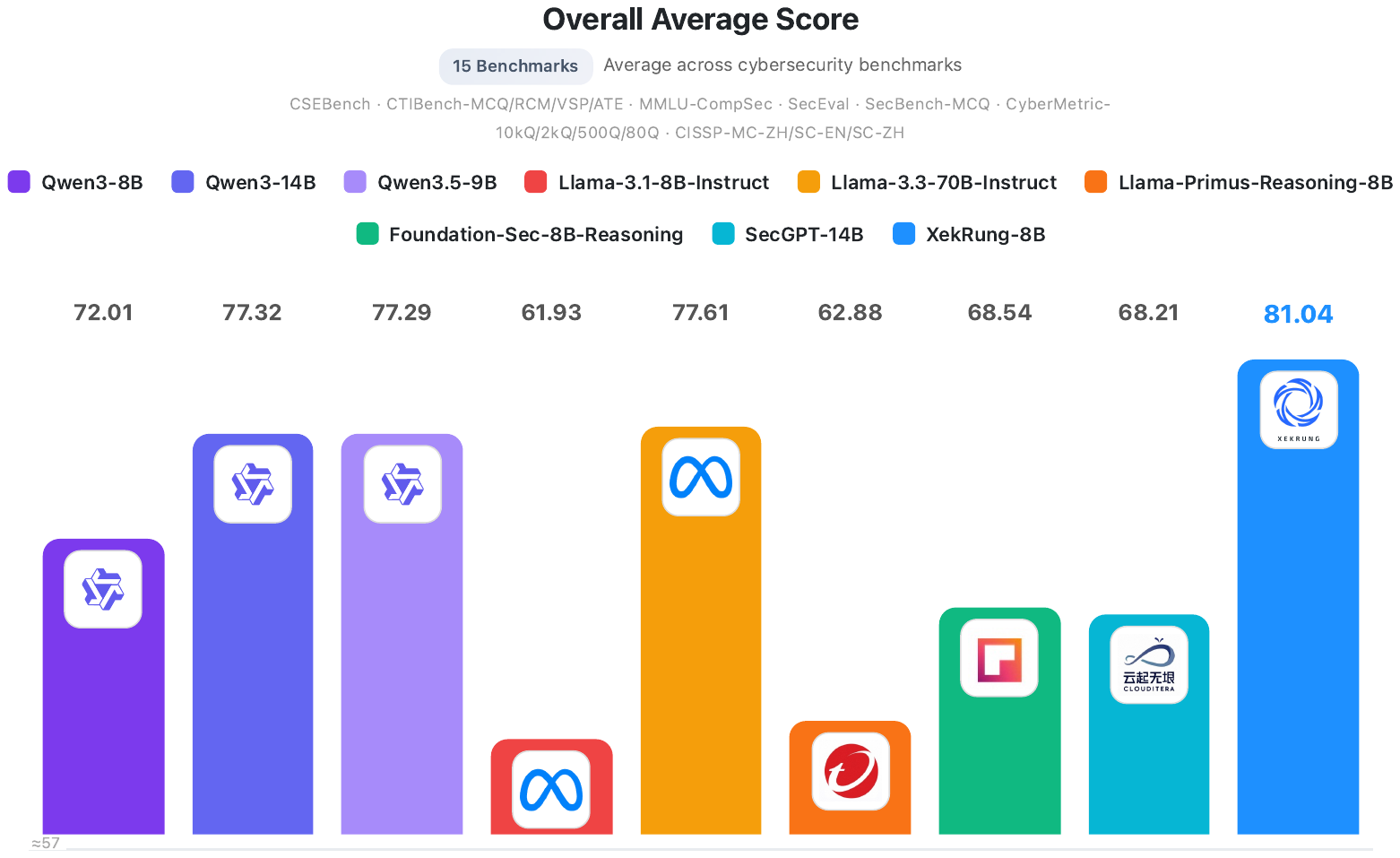}
    \caption{Overview of XekRung's performance across 15 cybersecurity benchmarks compared with general-purpose frontier models and existing security-specialized models. XekRung achieves leading results on all evaluated dimensions.}
    \label{fig:overview}
\end{figure}

\newpage

\section{Introduction}
\label{sec:intro}

The development of large language models (LLMs) has entered a pivotal phase. Model capabilities are evolving beyond elementary dialogue and factual recall toward assisting humans in tackling concrete, high-complexity real-world problems~\citep{grattafiori2024llama3,qwenteam2025qwen3,deepseek2024v3,jaech2024o1}. As reasoning and code generation proficiencies continue to mature, identifying application domains of sufficient difficulty has itself become a central proposition for probing and extending the capability frontier of LLMs.

Cybersecurity stands out as a particularly demanding domain, owing to a confluence of intrinsic technical challenges. First, it is inherently cross-domain and cross-layer, requiring simultaneous mastery of software engineering, operating systems, compiler internals, network protocols, and cryptographic primitives---disciplines whose interactions give rise to emergent complexity far exceeding any single constituent. Second, the associated search space is vast: adversarial behaviors can manifest across an enormous landscape of possible configurations, inputs, and execution paths, rendering exhaustive enumeration infeasible and demanding strategic, heuristic-guided exploration. Third, and most critically, causal chains are often deeply concealed and counterintuitive---a security vulnerability may stem from subtle interactions, such as a type confusion triggered under a rare compiler optimization path or a privilege escalation enabled by an undocumented kernel behavior, that demand multi-step reasoning to identify and resolve. Consider code vulnerability analysis as a representative case: diagnosing a vulnerability typically requires integrated knowledge spanning multiple system layers, coupled with the acuity to trace highly elusive causal chains through sprawling codebases---a task that epitomizes the multi-step, cross-domain reasoning challenge at the heart of cybersecurity.

Scaling laws remain potent today, with frontier general-purpose models advancing aggressively along the axes of data volume, data quality, and parameter count. Yet the aspiration of ``compression as intelligence'' has so far failed to yield a practically optimal compression ratio or strategy for vertical domains. General-purpose models---from the Llama series~\citep{grattafiori2024llama3} and the Qwen family~\citep{qwenteam2025qwen3} to DeepSeek-V4~\citep{deepseek2026v4} and GLM5~\citep{thudm2026glm5}---furnish broad linguistic and reasoning competencies, yet fall short of the demands of effective cybersecurity applications. The deficit extends well beyond a mere absence of domain factual knowledge---which might, in principle, be partially mitigated through retrieval-augmented context injection---to a deeper inability to internalize, reason over, and operationalize domain-specific concepts in the manner an expert practitioner would. Security tasks frequently require not just recalling that a particular vulnerability class exists, but understanding the precise conditions under which it manifests, reasoning about its exploitability across interacting system components, and synthesizing actionable remediation strategies grounded in operational constraints. This depth of domain comprehension---the capacity to fluidly integrate cross-layer knowledge, perform multi-step causal inference, and translate analytical conclusions into concrete security actions---cannot be reliably elicited through prompting or context augmentation alone; it necessitates the model having internalized the structural and causal patterns of the domain through dedicated training.

Most recently, frontier proprietary models have begun to turn their attention squarely to cybersecurity: Anthropic's Claude Mythos Preview~\citep{anthropic2026mythos} exhibits emergent offensive capabilities such as zero-day discovery and autonomous exploit chain construction, while OpenAI's GPT-5.4-Cyber~\citep{openai2026gpt54cyber} is fine-tuned specifically for defensive security analysis with relaxed refusal boundaries. These developments underscore the strategic importance the industry attaches to this domain, yet they remain ultra-large-scale, closed-source systems inaccessible for on-premise deployment. Compounding this, the cybersecurity domain is replete with data-sensitive scenarios: vulnerability intelligence, threat feeds, and incident logs frequently cannot be channeled to proprietary models through external APIs. These converging imperatives---the gap between surface-level knowledge and deep operational understanding, the practical mandate for local deployment, and the escalating capability bar set by proprietary systems---render it critically important to explore the optimal intelligence density at tractable model scales through targeted, efficient training of open-weight models.

Cybersecurity-specialized language models have proliferated rapidly in recent years. Foundation-Sec-8B~\citep{kassianik2025foundationsec8b} conducts continued pre-training atop Llama; Primus~\citep{yu2025primus} spans pre-training, instruction tuning, and reasoning distillation; and a host of additional efforts concentrate on SFT or alignment as isolated stages~\citep{jiang2024hackmentor,tihanyi2025cyberllminstruct,pentera2025securityllm}. Nevertheless, the preponderance of existing cybersecurity LLMs remain anchored to a narrow set of predefined tasks---such as phishing detection, malware classification, or specific CTI subtasks---via task-specific training on comparatively small curated datasets (typically 14K--55K samples). This circumscribed specialization engenders insufficient generalization across the breadth of the cybersecurity domain, which subsumes vastly diverse competencies ranging from vulnerability root-cause analysis and detection rule authoring to agentic penetration testing and incident response orchestration. Moreover, while the advent of native reasoning models such as OpenAI o1~\citep{jaech2024o1} and DeepSeek-R1~\citep{guo2025deepseekr1} has demonstrated the potency of reinforcement learning for complex problem-solving, the systematic application of RL across the full breadth of cybersecurity tasks---with their heterogeneous reward structures, diverse task formulations, and stringent operational constraints---remains largely underexplored. Existing efforts, where RL is employed at all, tend to target isolated subtasks rather than pursuing a unified, multi-task RL framework capable of jointly optimizing analytical reasoning, agentic workflows, and behavioral alignment within the security domain.

To address these challenges, we introduce \textbf{XekRung}, a cybersecurity-specialized large language model built upon the Qwen foundation~\citep{qwenteam2025qwen3}. XekRung undergoes a complete training pipeline spanning Continued Pre-Training (CPT), Supervised Fine-Tuning (SFT), and Multi-Task Reinforcement Learning, with each stage engineered around the unique data characteristics and task diversity of the cybersecurity domain. Our methodology is predicated on extreme capability decomposition and compositional generalization: we systematically disentangle the core task dimensions of cybersecurity---vulnerability analysis, threat intelligence, secure code engineering, and agentic security operations---and reassemble them through targeted training that engenders mutual reinforcement across dimensions.

To bridge the gap between surface-level knowledge and deep operational understanding (the core limitation of general-purpose models identified above), we devise domain-aware data construction and synthesis strategies that enrich rare yet decision-critical security concepts, model cross-artifact dependencies, and transform opaque security artifacts into self-explanatory training instances---collectively enabling the model to internalize the structural and causal patterns of the domain rather than merely memorizing isolated facts. To overcome the limitations of narrow task-specific RL (the key deficit of existing domain models), we design a unified multi-task RL framework that integrates Reasoning RL with verifiable rewards for analytical tasks and Agentic RL for multi-step tool-augmented operational workflows, complemented by an adversarial self-evolution mechanism for continuous capability escalation. The current release, XekRung, is architected as a non-thinking model that directly emits high-fidelity responses without explicit intermediate reasoning traces, delivering state-of-the-art results with exceptional token efficiency---a decisive advantage for latency-sensitive security operations such as real-time alert triage and interactive threat analysis. To accommodate heterogeneous computational budgets, we additionally devise a knowledge distillation pipeline that transfers aligned capabilities into compact variants ($\leq$4B).

As depicted in Figure~\ref{fig:overview}, we carry out extensive evaluation on a multi-dimensional benchmark suite comprising 35 general-purpose benchmarks across 9 categories and 15 cybersecurity-specific benchmarks spanning threat intelligence, vulnerability analysis, security knowledge, and certification. XekRung-8B achieves an overall average of $81.04\%$ across all 15 security benchmarks, representing a $+9.03$ percentage point improvement over its base model Qwen3-8B ($72.01\%$) and surpassing all same-scale baselines by a substantial margin---including Foundation-Sec-8B-Reasoning ($68.54\%$) and Llama-Primus-Reasoning-8B ($62.88\%$). Notably, XekRung-8B outperforms Llama-3.3-70B-Instruct ($77.61\%$), a model with $8.75\times$ more parameters, and exceeds SecGPT-14B ($68.21\%$) by over $12$ percentage points on domain tasks. Crucially, these domain gains come with minimal degradation of general capabilities: XekRung-8B retains $72.54\%$ overall on general-purpose benchmarks, closely tracking the base Qwen3-8B ($73.31\%$). These findings corroborate that the synergy of a well-integrated, full-pipeline training paradigm, domain-tailored data engineering, and multi-task RL can unlock performance that markedly transcends narrow task-specific approaches in the demanding cybersecurity vertical.

Our core contributions are as follows:

\paragraph{Domain Data Technology.}
We build a data engineering system spanning the entire training lifecycle, combining learned quality scoring, capability-level interference balancing, and vector-space deduplication on the quality side, with contextual enrichment, cross-artifact aggregation, and targeted synthesis pipelines on the construction side---collectively expanding the frontier of high-quality cybersecurity training data.

\paragraph{Complete and Innovative Training Pipeline.}
XekRung realizes a full CPT $\rightarrow$ SFT $\rightarrow$ RL pipeline with task-aware specialization for cybersecurity. Post-training unifies Reinforcement Learning with Verifiable Rewards (RLVR) for verifiable-reward tasks and Agentic RL for end-to-end workflow optimization, complemented by code-verified adversarial self-evolution and vulnerability self-play.

\paragraph{Comprehensive Evaluation.}
We systematize evaluation into a multi-dimensional protocol covering 35 general and 15 cybersecurity-specific benchmarks, jointly assessing domain proficiency and general-capability retention to resist narrow-task overfitting. XekRung-8B achieves state-of-the-art security performance among same-scale models while competing with several-times-larger models in non-thinking mode with exceptional token efficiency.

The remainder of this report is organized as follows. Section ~\ref{sec:related_work} surveys related work on general-purpose and cybersecurity-specialized LLMs. Section ~\ref{sec:pretrain} elaborates the pre-training strategy, encompassing data construction, domain-specific data strategies, and mid-training. Section ~\ref{sec:post_train} delineates the post-training pipeline, including SFT, multi-task RL, self-evolution mechanisms, and knowledge distillation. Section ~\ref{sec:experiments} presents comprehensive evaluation results and analysis. Section ~\ref{sec:conclusion} concludes the report and charts future directions.

\section{Related Work} \label{sec:related_work}
We situate our work within two closely related research directions:
general-purpose large language models (LLMs) and cybersecurity-specialized LLMs.
For comprehensive overviews of LLMs in cybersecurity, we refer readers to recent
surveys~\citep{xu2024llmcybersecurity, zhang2024llmcybersecurity}.

\subsection{General Large Language Models} \label{sec:related_llm}
The rapid advancement of open-weight LLMs has laid a strong foundation for domain specialization. Meta's Llama series~\citep{grattafiori2024llama3}, trained on over 15 trillion tokens, has become the popular base model for domain adaptation, with Llama-3.1-8B-Base in particular demonstrating robust capabilities in reasoning and code generation that make it a natural initialization point for security-focused models. Alibaba's Qwen series~\citep{bai2023qwen,yang2024qwen2,qwenteam2025qwen3,qwenteam2025qwen35} offers a compelling alternative spanning parameter scales from 0.5B to 397B, combining strong Chinese--English bilingual performance with competitive coding abilities that benefit security applications requiring cross-lingual transfer and complex code understanding. Other notable architectures, including DeepSeek-V3~\citep{deepseek2024v3} and GLM-5~\citep{thudm2026glm5}, leverage Mixture-of-Experts (MoE) designs to achieve high throughput with efficient inference. More recently, proprietary frontier models such as Claude Mythos~\citep{anthropic2026mythos} and GPT-5.4-Cyber~\citep{openai2026gpt54cyber} have raised the capability bar substantially through massive parameter scale and access to proprietary data, establishing increasingly demanding upper bounds for security-relevant reasoning. Against this backdrop, a parallel line of work has focused on instilling structured \emph{reasoning} into language models. OpenAI o1~\citep{jaech2024o1} pioneered explicit chain-of-thought traces at inference time, while DeepSeek-R1~\citep{guo2025deepseekr1} introduced Group Relative Policy Optimization (GRPO)~\citep{shao2024grpo}, which replaces a learned critic with intra-group relative rewards and allows deliberate reasoning behavior to emerge organically during reinforcement learning. Our work adopts GRPO as the primary RL algorithm, though XekRung is deliberately designed as a \textit{non-thinking model} that directly outputs high-fidelity operational sequences without explicit intermediate reasoning steps---pursuing an orthogonal trajectory to frontier scale by maximizing domain intelligence density at tractable model sizes.

\subsection{Cybersecurity-Specialized Language Models} \label{sec:related_security_llm}
A growing body of work has adapted LLMs specifically for cybersecurity since 2024, as summarized in Table~\ref{tab:security_llm_comparison}. Early efforts concentrated on injecting domain knowledge through continued pretraining (CPT): Foundation-Sec-8B~\citep{kassianik2025foundationsec8b} and Primus~\citep{yu2025primus} demonstrate that large-scale CPT on security corpora substantially improves domain understanding, yet both still rely on generic instruction tuning or distillation for downstream tasks, which can limit their capacity to execute complex, multi-step operational workflows. Recognizing this gap, a parallel line of work turned to supervised fine-tuning (SFT) on curated instruction datasets. Models such as Lily-Cybersecurity-7B~\citep{segolilylabs2025lily}, HackMentor~\citep{jiang2024hackmentor}, CyberLLMInstruct~\citep{tihanyi2025cyberllminstruct}, and Colibri~\citep{colibria2025colibri8b} show that task-specific supervision can yield strong performance on well-defined problems such as phishing analysis or vulnerability classification, though the relatively small data scales involved (typically 14K--55K samples) and the absence of deep knowledge injection often constrain generalization. More recent models have pushed further by appending reinforcement learning stages: BaronLLM~\citep{baronsecurity2025baronllm}, SecurityLLM~\citep{pentera2025securityllm}, and Foundation-Sec-8B-Reasoning~\citep{yang2026foundationsec8breasoning} each combine SFT with RL to improve alignment and reasoning reliability, though none begins from a domain-pretrained checkpoint, leaving a gap in foundational domain coverage. SecGPT-2.0~\citep{secgpt} is the closest predecessor to our own pipeline, employing CPT followed by SFT and RL on Qwen2.5-14B. XekRung builds on this direction by instantiating the same three-stage paradigm---CPT on 5 billion tokens of curated cybersecurity data, multi-task SFT for broad task generalization, and GRPO-based reinforcement learning for reliable complex reasoning---on the more capable Qwen3-8B backbone, while pursuing a pragmatic balance between domain performance and deployment cost that large proprietary models cannot easily achieve.

\begin{table*}[t]
\centering
\caption{Comparison of representative cybersecurity LLMs by training pipeline.}
\label{tab:security_llm_comparison}
\begin{tabular}{@{}lcc@{}}
\toprule
\textbf{Model} & \textbf{Base Model} & \textbf{Training Pipeline} \\
\midrule
Foundation-Sec-8B~\citep{kassianik2025foundationsec8b} & Llama3.1-8B & CPT \\
Primus~\citep{yu2025primus} & Llama3.1-8B & CPT $\rightarrow$ SFT \\
Lily-7B~\citep{segolilylabs2025lily} & Mistral-7B-v0.2 & SFT \\
HackMentor~\citep{jiang2024hackmentor} & Llama/Vicuna & SFT \\
CyberLLMInstruct~\citep{tihanyi2025cyberllminstruct} & — & SFT \\
Colibri~\citep{colibria2025colibri8b} & Llama3-8B & SFT \\
BaronLLM~\citep{baronsecurity2025baronllm} & Llama3.1-8B & SFT $\rightarrow$ RL \\
SecurityLLM~\citep{pentera2025securityllm} & Llama3.1-8B & SFT $\rightarrow$ RL \\
Foundation-Sec-8B-Reasoning~\citep{yang2026foundationsec8breasoning} & Llama3.1-8B & SFT $\rightarrow$ RL \\
SecGPT-2.0~\citep{secgpt} & Qwen2.5-14B & CPT $\rightarrow$ SFT $\rightarrow$ RL \\
\textbf{XekRung (Ours)} & Qwen3-8B & CPT $\rightarrow$ SFT $\rightarrow$ RL \\
\bottomrule
\end{tabular}
\end{table*}

\section{Pre-training}
\label{sec:pretrain}

This section presents the pre-training strategy of XekRung. The corpus consists of heterogeneous cybersecurity data from multiple sources, curated through a unified pipeline of cleaning, augmentation, and format normalization. On top of the curated corpus, we introduce code-centric signals that explicitly encode engineering experience, thereby improving the efficiency with which tacit domain expertise is absorbed. Our training curriculum further separates base pre-training from a dedicated mid-training stage, in which higher-order capabilities---long-horizon agentic planning, supervised task knowledge, and repository-level long-context code understanding---are injected without compromising the general competence established in the base stage.

\subsection{Data Curation and Processing}
\label{sec:data_curation}

Building a cybersecurity-specialized foundation model requires a training corpus that captures both the breadth and depth of the domain, ranging from high-level threat intelligence narratives to low-level binary analysis artifacts. We organize the data pipeline into four stages: collection, processing, domain-specific data strategies, and data mixture.

\subsubsection{Data Collection}
\label{sec:data_collection}

We assemble a large-scale, multi-source cybersecurity corpus that spans five complementary categories:

\begin{itemize}
 \item \textbf{Open-source and external data.} Publicly available security content collected from the web, technical write-ups from researchers and vendors, academic publications, professional forums, educational materials on systems security and cryptography, and curated open corpora.
 \item \textbf{Internal product data.} Structured and semi-structured records from deployed security systems, including detection rule descriptions, alert metadata, threat intelligence entries, and product-generated analytical reports.
 \item \textbf{Internal knowledge documents.} Expert-authored technical materials covering detection methodologies, incident response playbooks, and domain knowledge bases maintained by security analysts.
 \item \textbf{Log data.} Real-world security telemetry at the network, endpoint, and application layers, exposing the model to authentic operational patterns and anomalous behaviors.
 \item \textbf{Code data.} Security-relevant source code spanning proof-of-concept exploits, detection rules, security tool implementations, and large-scale repositories with CVE-annotated patches.
\end{itemize}

We construct a security knowledge system by combining both top-down and bottom-up approaches. On the one hand, we draw on industry standards and expert-curated security knowledge frameworks. On the other hand, we automatically extract knowledge from real-world operational logs to build a knowledge graph. By continuously comparing and refining these two sources, we connect and improve them over time, ultimately forming a security knowledge system that is more practical and better aligned with real-world scenarios.

Consequently, the resulting system effectively unifies industry-general security frameworks with organization-specific operational policies. Taking vulnerability management as a concrete example, XekRung incorporates foundational knowledge (e.g., vulnerability definitions, exploitation mechanisms, attack surfaces, and remediation strategies) while seamlessly integrating usage documentation for internal vulnerability scanners. This ensures full-lifecycle coverage from initial discovery to security advisory publication. Furthermore, expert experiential knowledge is distilled into chain-of-thought (CoT) training data, with all resultant samples undergoing rigorous quality review to construct a high-signal corpus essential for real-world vulnerability identification.

\subsubsection{General Data Processing}
\label{sec:data_processing}

We adhere to established best practices for large-scale pre-training data curation~\citep{grattafiori2024llama3,kassianik2025foundationsec8b}, implementing a rigorous multi-stage processing pipeline:

\begin{itemize}
 \item \textbf{Deduplication.} We perform both exact deduplication (document-level hashing) and fuzzy deduplication (MinHash-based near-duplicate removal) within and across data sources to eliminate redundancy.
 \item \textbf{Quality filtering.} Heuristic filters (document length, language identification, perplexity scoring) are combined with learned quality classifiers trained on expert-curated positive and negative examples to exclude low-quality, off-topic, or uninformative documents.
 \item \textbf{Sensitive information removal.} We systematically strip personally identifiable information (PII; e.g., names, email addresses, phone numbers), anonymize internal identifiers and proprietary asset references, and selectively truncate or remove long, semantically vacuous encoded strings (e.g., Base64 blocks, URL tracking parameters, session tokens). This mitigates privacy risks and reduces noise without sacrificing actionable security content.
 \item \textbf{Format normalization.} Heterogeneous inputs (e.g., HTML, PDF, Markdown, structured logs) are normalized into a unified plain-text representation with consistent whitespace and encoding conventions. Crucially, code blocks and structured fields are preserved intact where semantically relevant.
\end{itemize}

\subsubsection{Domain-Specific Data Strategies}
\label{sec:domain_strategies}

While the unified pipeline resolves generic data quality issues, cybersecurity corpora present several domain-specific challenges: critical artifact types suffer from severe data sparsity; 
semantically related threat intelligence is fragmented across disparate sources; and genuine long-range dependencies are rarely captured by naive document concatenation. To address these, we introduce four targeted strategies to augment the corpus along the dimensions of representation, 
correlation, long-range dependency and engineering experience injection.

\paragraph{Explanatory Annotation for Domain-Specific Artifacts.}

General-purpose pre-training corpora severely underrepresent artifact types central to practical cybersecurity---such as disassembly and decompiler outputs (e.g., IDA~Pro/Ghidra pseudocode), raw shellcode and ROP gadget chains, packet capture (PCAP) dissections, detection rule languages (Suricata, Snort, YARA), kernel-level tracing (eBPF), debugger transcripts, and fuzzer crash reports. In their raw form, these artifacts manifest as domain-specific sequences of memory addresses, hexadecimal bytes, or domain-specific language (DSL) tokens, rendering direct representation learning highly inefficient. To overcome this, we pair each raw artifact with a concise, synthetically generated natural-language explanation that articulates its semantics, security implications, and operational context, effectively transforming specialized tool outputs into self-contained, semantically rich training instances.

\paragraph{Correlated Information Aggregation.}
\label{sec:correlated_aggregation}
Semantically linked security data, ranging from CVE descriptions and PoC code to threat intelligence reports, are inherently dispersed across heterogeneous repositories. Conventional pre-training pipelines, which rely on random document concatenation, fail to capture these crucial dependencies, severely limiting a model's capacity for end-to-end cybersecurity incident correlation.
To bridge this gap, we employ correlated information aggregation strategy. Instead of random sampling, we construct unified training sequences by aggregating related corpus around shared semantic anchors (e.g., CVE identifiers, software components). This aggregation is executed via a three-stage, coarse-to-fine pipeline designed to optimize information flow. First, we employ hierarchical semantic clustering to partition documents, moving from broad security domains down to specific incident-level groupings based on dense embedding similarity. Second, within these fine-grained groups, we perform complementarity-aware context construction, assembling training windows that balance semantic coherence with artifact type diversity to enrich the context.
Finally, we reorder documents within each sequence by their informativeness. We estimate how informative each document is and place the most informative ones at the beginning of the sequence. This takes advantage of a well-known property of Transformer attention: tokens at the start of a sequence tend to receive more attention from later tokens, a phenomenon often referred to as primacy bias~\citep{yen2024helmet}. By front-loading the most informative content, we ensure that the model attends to the most important context when processing the rest of the sequence. This sequence-level reordering is orthogonal to domain-level data mixing methods such as DoReMi~\citep{xie2023doremi}, and the two can be combined.

\paragraph{Long-Context Data Synthesis via Entropy.}
\label{sec:longctx_synth}

Constructing training data with genuine long-range dependencies remains a critical bottleneck, as naive document concatenation fails to guarantee effective cross-span interactions. Building upon EntropyLong~\citep{jia2025entropylong}, we leverage the 
entropy
to guide data synthesis. Specifically, we identify high-entropy positions in source documents as information gaps and retrieve semantically relevant candidate passages from an external corpus. Through a model-in-the-loop verification process, we filter out spurious dependencies by retaining only candidates that significantly reduce the prediction entropy at these anchor positions. The verified passages are then shuffled and prepended to the source document. This uncertainty-driven approach ensures quantifiable information gain for synthesized dependencies and prevents the exploitation of superficial positional patterns. Crucially, the fundamental long-range capabilities established during this phase lay the structural foundation within the attention mechanism for processing genuine, long-horizon agent trajectories during mid-training (see \S~\ref{sec:midtrain}).

\paragraph{Explicit Orchestration of Engineering Experience.}
While the aforementioned data curation pipeline ensures high-quality textual representations, a central challenge remains: much of the domain's highest-value knowledge is rarely expressed as declarative text; instead, it is latent in the evolutionary history of real-world codebases. To bridge this representational gap, we construct training data that explicitly surfaces such latent knowledge, following a three-tier progressive design: static code differentials, dynamic commit trajectories, and iterative self-correction traces.

\begin{itemize}
 \item \textbf{Differential Code Learning.} Rather than training on isolated code snapshots, each training sample pairs vulnerable code with its corresponding patch and remediated output. This formulation compels the model to reason about why a modification is necessary and to infer the semantic intent underlying the change, moving beyond mere patch prediction.
 \item \textbf{Commit Trajectory Modeling.} Chronologically ordered commit sequences extracted from version-control histories are organized into trajectory samples. This enables the model to internalize the end-to-end reasoning chain---spanning fault discovery, root-cause localization, and patch generation---that epitomizes expert software engineering practice. This approach aligns with recent findings in reinforcement learning from software evolution~\citep{wei2025swerl}, which demonstrate that trajectory-structured signals yield substantially richer supervision than static code corpora.
 \item \textbf{Iterative Self-Correction Traces.} In real-world engineering workflows, patches undergo successive revisions, review comments are addressed iteratively, and solutions converge through continuous refinement. We incorporate samples that capture this inherently iterative structure, thereby providing explicit supervision for error recovery---a capability indispensable for reliable automated vulnerability remediation and code review.
\end{itemize}

\subsubsection{Data Mixture}
\label{sec:data_mixing}

The data mixture is critical to pre-training outcomes: over-emphasizing domain-specific data risks eroding general capabilities, whereas under-representing it limits domain depth. Rather than relying on any single heuristic, we conduct systematic ablations along two complementary axes.

\begin{itemize}
 \item \textbf{Data Type.} We vary the proportions of code, mathematical reasoning data, general text, and security-specific text, and jointly evaluate each configuration on both cybersecurity and general-purpose benchmarks to identify an effective trade-off.
 \item \textbf{Data Provenance.} We balance internal sources, which provide operational depth and specificity, against external sources, which contribute broader coverage and expressional diversity.
\end{itemize}

The final mixture is selected by jointly considering domain performance, general-capability retention, and training stability rather than any single metric. Overall, we follow the mixing principle of preserving general competence while strengthening cybersecurity expertise, enabling the model to substantially absorb domain knowledge without materially sacrificing its general understanding and reasoning ability. It is important to note that this mixture configuration primarily governs the base pre-training phase. For highly structured and specialized data formats---such as code evolution traces and agentic trajectories---we implement stage-wise, dynamic mixing schedules, which are detailed in 
\S~\ref{sec:midtrain}.

\subsection{Mid-Training}
\label{sec:midtrain}

Base pre-training establishes a broad foundation of security knowledge and language comprehension. However, eliciting advanced capabilities---such as long-horizon planning, multi-step tool use, and repository-scale code reasoning---requires targeted exposure to specialized data distributions and formats. Injecting these formats too early disrupts the convergence of the primary language modeling objective; conversely, introducing them post-crystallization of base capabilities yields suboptimal absorption. Consequently, we design a dedicated mid-training stage between continued pre-training and supervised fine-tuning to concentrate the injection of these higher-order capabilities without compromising previously established generalization.

Concretely, mid-training serves two complementary purposes: (i) injecting diverse task knowledge into the model before the SFT format is imposed, and (ii) establishing long-horizon reasoning capacity prior to post-training. Around these two purposes, we incorporate four data and scheduling strategies: (1) long-horizon agentic trajectories for multi-step planning; (2) repository-level long-context code samples that support cross-file security reasoning; and (3) linearized SFT data that transfers supervised knowledge while preserving representational flexibility.

\subsubsection{Long-Horizon Agentic Data Injection}
\label{sec:agentic_midtrain}

Complex security agent tasks---automated penetration testing, large-scale vulnerability detection and validation, red-team long-chain attack simulation, and multi-stage incident response---require the model to maintain goal consistency, state tracking, and strategy adjustment across extended procedural chains. Learning such capabilities presupposes broad world knowledge, after which the model must additionally acquire the ability to manage long-horizon procedures. Moreover, the distinctive observation--thought--action (O-T-A) format of agent trajectories, if introduced too early, risks interfering with the convergence of the generic language modeling objective. Following a breadth-first, specialization-second curriculum principle, we therefore reserve long-horizon agentic data for injection during mid-training.

\paragraph{Two-Dimensional Curriculum Learning.} 
To inject agentic capabilities efficiently without disrupting general competencies, we design a two-dimensional curriculum learning strategy encompassing both a dynamic mixing schedule and progressive length scaling. 
Along the proportion dimension, rather than using a fixed ratio, we ramp up agentic data dynamically as base competencies consolidate. Let $T_{\text{mid}}$ denote the total number of mid-training steps; at step $t$, the mixing weight for agentic data is
\begin{equation}
\alpha_{\text{agent}}(t) \;=\; \alpha_{\min} + (\alpha_{\max} - \alpha_{\min}) \cdot \sigma\!\left(\frac{t - T_{\text{mid}}/2}{\tau_{\text{warm}}}\right),
\label{eq:agent-schedule}
\end{equation}
where $\sigma(\cdot)$ is the sigmoid function and $\tau_{\text{warm}}$ controls the warm-up rate. This schedule instantiates a breadth-first curriculum by gradually shifting the agentic data proportion from a low initial value to a higher terminal value, preventing premature saturation by specialized content before foundational knowledge is fully absorbed. 

Along the length dimension, agentic tasks - such as multi-step penetration testing and interactive debugging---produce trajectories spanning tens of thousands of tokens. Training on full-length sequences from the outset induces unstable gradients and poor convergence. We therefore adopt a short-to-long progressive schedule: training starts with shorter trajectory segments, and the sequence length is gradually increased over the course of mid-training. This curriculum allows the model to first consolidate local reasoning patterns (e.g., a single tool invocation and its observation) before learning to maintain coherence and state tracking across extended multi-turn interactions. This joint curriculum ensures that by the end of mid-training, the model reliably processes full-length agentic trajectories, providing a robust initialization for post-training agentic fine-tuning.

\subsubsection{Repository-Level Long-Context Code Understanding}
\label{sec:repo_longctx}

Real-world vulnerability localization and repair typically span multiple files and require holistic reasoning over inter-module dependencies rather than the processing of isolated snippets. We therefore incorporate repository-level samples that expose the model to complete project contexts---including cross-file call relations, module dependencies, and interface contracts---in order to bridge the gap between snippet-level pre-training capability and the cross-file reasoning demanded by production security tasks such as enterprise codebase auditing and multi-component vulnerability analysis. Combined with the progressive long-context schedule described above for agentic trajectories, repository-level samples further consolidate the model's ability to maintain structural understanding across long sequences, laying the foundation for code-agent and security-audit capabilities acquired in post-training.

\subsubsection{Linearized SFT Data Integration}
\label{sec:linearized_sft}

We integrate a diverse collection of SFT datasets---spanning cybersecurity question answering, code understanding, vulnerability analysis, and general instruction-following---by stripping chat templates and role markers, reformulating each sample as a plain-text continuation sequence, and training it under the same autoregressive language modeling objective used throughout pre-training. This linearization allows the model to internalize the knowledge and reasoning patterns carried by high-quality supervised data while avoiding premature commitment to rigid instruction--response formatting, thereby preserving representational flexibility for the subsequent SFT stage.
\section{Post-Training}\label{sec:post_train}

This section provides a comprehensive account of the post-training technical framework. The pipeline systematically integrates rigorous data quality control, scalable synthetic data generation, and advanced alignment methodologies, encompassing supervised fine-tuning (SFT), multi-task reinforcement learning (RL), and adversarial self-evolution mechanisms, as well as structured knowledge distillation that transfers the aligned capabilities of large-scale models into compact models for efficient deployment.

\subsection{Post-Training Pipeline}
\label{sec:post_train_pipeline}

Addressing the core challenges of post-training---namely, rigorous capability alignment, specialized security knowledge acquisition, and the mitigation of cross-task interference---our methodology is architected as a progressively optimized pipeline. To maximize model capabilities across diverse computational constraints, we decouple the overarching system into a multi-stage training paradigm for large-scale foundational models, complemented by a structured knowledge distillation mechanism specifically tailored for compact deployment scenarios.

\begin{figure}[t]
    \centering
    \includegraphics[
        width=\linewidth
    ]{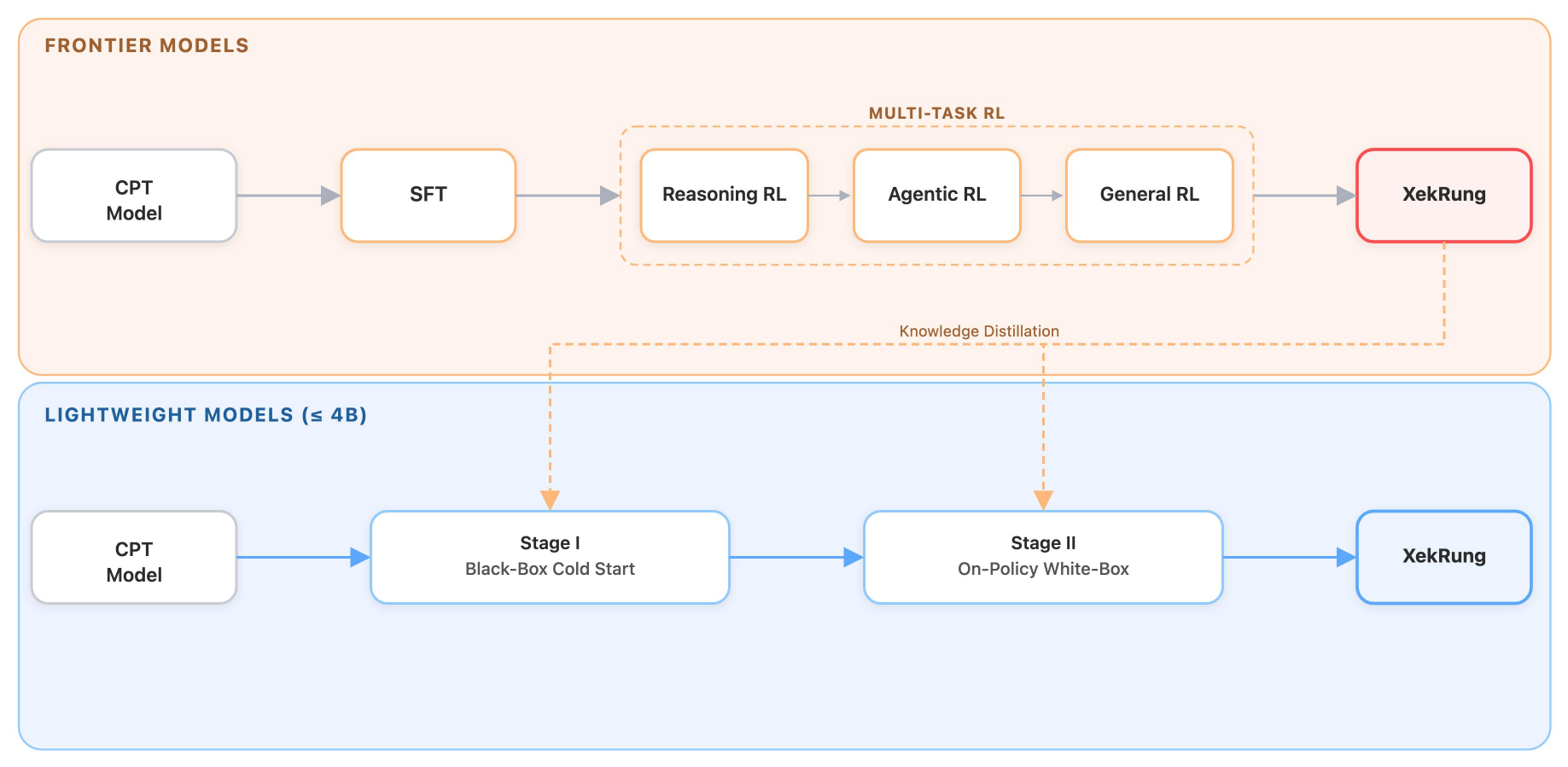}
    \caption{Overview of the post-training pipeline. The framework comprises two complementary tracks: the \textsc{FRONTIER MODELS} track (\S~\ref{sec:multi_stage_training}, Multi-Stage Training for Large-Scale Models) and the \textsc{LIGHTWEIGHT MODELS} track (\S~\ref{sec:distillation}, Knowledge Distillation for Compact Models).}
    \label{fig:post_train_pipeline}
\end{figure}

The core framework is operationalized through three progressive stages: SFT, multi-task RL, and self-evolutionary refinement. For each stage, we formulate the explicit training objectives---detailing relevant reward functions and loss expressions---before elucidating the underlying data composition and generic preprocessing protocols. Subsequent subsections delve into the specific techniques employed to optimize data utility and augment model proficiency across reasoning, code synthesis, and agentic cybersecurity tasks. Figure~\ref{fig:post_train_pipeline} illustrates the overall architecture of our post-training pipeline. We elaborate on each track in the following sections.

\subsubsection{Multi-Stage Training for Large-Scale Models}
\label{sec:multi_stage_training}

For large-scale backbone models, the post-training trajectory transitions rigorously from behavioral cloning to reinforced exploration. The model initially internalizes domain-specific structural priors and logic via SFT. Building upon this foundation, it advances to a multi-task RL phase powered by Group Relative Policy Optimization (GRPO). Ultimately, a self-evolutionary loop grounded in verifiable code execution ensures sustained breakthroughs in complex reasoning.

\subsubsection{Knowledge Distillation for Compact Models}
\label{sec:distillation}

Operating downstream of the large-scale training paradigm, we develop a structured knowledge distillation pipeline for compact deployment. Ultimately, rather than expending substantial computational resources to cycle compact architectures through the full post-training pipeline independently, we systematically distill the aligned capabilities of our large-scale models into lightweight variants ($\leq$4B) via a structured two-stage paradigm.

\textbf{Stage I: Black-Box Cold Start.}\quad
The process commences with the teacher model synthesizing high-fidelity target sequences in response to diverse, meticulously curated query clusters. The student model then undergoes explicit SFT against these sequences. By bypassing the prohibitive memory overhead of probing internal teacher representations, this mechanism efficiently repositions the student's parameter weights into an optimization basin primed for subsequent intensive logit-level supervision.

\textbf{Stage II: On-Policy White-Box Distillation.}\quad
While Stage I effectively establishes a behavioral prior, it fundamentally leaves the student model unexposed to its own autoregressive error propagation during autonomous inference. To resolve this distribution shift, we pivot to a rigorous on-policy generation scheme~\citep{agarwal2024policy}. Herein, the student model continuously generates responses online, while the teacher concurrently furnishes localized, token-level logit supervision to rectify the student's actualized trajectories. This compels the student to actively diagnose and correct its native generation patterns, yielding robust improvements in out-of-distribution generalization.

For the explicit distillation objective, we supplant traditional KL divergence with ABKD~\citep{wang2025abkd}, a state-of-the-art framework predicated upon parameterized $\alpha$-$\beta$-divergence:
\begin{align}
\mathbb{D}_{\mathrm{AB}}^{(\alpha,\beta)}(p \,\|\, q)
= -\frac{1}{\alpha\beta}\sum_{k}\left[
p_k^{\alpha}q_k^{\beta}
- \frac{\alpha}{\alpha+\beta}p_k^{\alpha+\beta}
- \frac{\beta}{\alpha+\beta}q_k^{\alpha+\beta}
\right]
\end{align}

where $p$ and $q$ denote the teacher and student distributions, respectively. Integrating dual free parameters, this formulation fluently spans both standard forward and reverse KLD, explicitly optimizing probability mass transfers to account for exact teacher-student capacity discrepancies. 

This formulation is further augmented via top-$K$ selective distillation, restricting the divergence loss computation strictly to the highly consolidated top-$K$ indices of the teacher's output distribution. Consequently, this prevents the student from wasting its constrained representational capacity attempting to fit fundamentally stochastic long-tail noise. Finally, employing distillation-aware model merging~\citep{wortsman2022model,yadav2024ties}, we seamlessly hybridize distinct checkpoints demonstrating complementary strengths (e.g., analytical reasoning versus instruction adherence), realizing targeted enhancements without additional computational expenditure.

\subsection{Post-Training Data}
\label{sec:post_training_data}

The efficacy of the post-training pipeline is fundamentally contingent upon the design of its underlying data. This subsection systematically addresses the key data-level decisions across three dimensions: we first outline the data configuration for each training stage (\S\ref{sec:data_stage_overview}), then present the diversity and synthesis strategies adopted to broaden distributional coverage (\S\ref{sec:data_diversity_synthesis}), and finally detail the quality control mechanisms that safeguard sample fidelity and mitigate multi-task interference (\S\ref{sec:data_quality_control}).

\subsubsection{Overview of Post-Training Data by Stage}
\label{sec:data_stage_overview}

We begin with an overview of the data configuration across training stages. Designed to furnish customized supervisory signals, the SFT dataset comprises approximately 300k instances strictly curated to enforce foundational instruction adherence. For the subsequent multi-task RL phase, the corpus is structurally distilled into roughly 120k high-complexity prompts---allocated across 70\% security-specific and 30\% general reasoning tasks---explicitly provisioned to support exhaustive online exploration and objective reward verification. The efficacy of these empirical phases, however, is fundamentally dictated by the inherent quality and diversity of these underlying corpora, which we address from the perspectives of data diversity and quality control in the following \S\ref{sec:data_diversity_synthesis} and \S\ref{sec:data_quality_control}.

\subsubsection{Data Diversity and Scalable Synthesis}
\label{sec:data_diversity_synthesis}

We first address data diversity. Specialized domains such as cybersecurity remain intrinsically bottlenecked by labeled data scarcity, compounded by prohibitive expert annotation costs and the sensitivity of authentic vulnerability samples. To circumnavigate these limitations, we engineered scalable synthetic data pipelines encompassing knowledge-grounded generation, long-context formulation, agentic trajectory synthesis, and repository context recovery.

\paragraph{Knowledge-Grounded Synthesis and Cross-Validation.}
For foundational SFT data, we adopt a knowledge-grounded synthesis paradigm~\citep{wang2023selfinstruct}. By utilizing structured security documentation as explicit conditioning context, we drive an LLM to generate robust question-answer pairs. Anchoring the generative process in authoritative references effectively neutralizes the factual hallucinations pervasive in unconstrained synthesis. To rigorously guarantee answer validity, a cross-validation filtering mechanism is applied: a secondary model answers each generated prompt independently, and the instance is retained exclusively if both outputs demonstrate strict semantic consistency, thereby systematically purging reasoning anomalies.

\paragraph{Long-Context Synthesis.}
To augment the model's proficiency in parsing and reasoning over prolonged inputs, we construct extensive context instances utilizing NExtLong~\citep{gao2025nextlong}. This framework synthesizes robust long-context data relies exclusively on short source documents by: (i) decomposing the document into semantically coherent meta-chunks; (ii) retrieving topically related hard-negative distractors from pre-training corpora; and (iii) interleaving these distractors between meta-chunks to forcefully expand sequence length. This compels the model to isolate authentic dependencies amid substantial noise.

Parallel to this, constructing instances that exhibit genuinely verifiable long-range dependencies remains a critical challenge. Addressing this, we adapt the core principles of EntropyLong~\citep{jia2025entropylong}, employing the model's intrinsic predictive uncertainty (Shannon entropy) to guide a four-stage pipeline: (1) executing a forward pass to locate information-gap anchors where prediction entropy exceeds a dynamic threshold; (2) retrieving semantically relevant candidate passages using the anchor's local context; (3) deploying a model-in-the-loop verification step that retains only candidates proven to reduce anchor-position entropy by over 40\%; and (4) randomly shuffling and appending these verified passages to the source document. This ensures every synthesized dependency confers quantifiable informational gain.

\paragraph{Agent Trajectory Synthesis for CTF.}
Capture-the-Flag (CTF) challenges present profound obstacles for RL methodologies, as their ephemeral execution environments (e.g., Docker sandboxes) render standard online rollouts practically infeasible. To solve this environmental bottleneck, we implement a writeup-driven trajectory synthesis framework inspired by Cyber-Zero~\citep{zhuo2025cyberzero}. Two agents interact via a structured protocol: a player model endeavors to solve the challenge from first principles, while a Terminal model simulates environmental feedback against the reference solution. When the player stalls, the Terminal injects minimal diagnostic guidance. Crucially, the player is subsequently mandated to diagnose its failure and execute self-correction as an internally derived insight. This selective intervention preserves authentic recovery dynamics, yielding training signals that encapsulate the critical cognitive process of failure, reflection, and correction. 

\paragraph{Repo Context Recovery.}
A substantial proportion of existing vulnerability datasets are confined to isolated snippet-level annotations (e.g., solitary functions). While preserving localized critical code, they systematically omit the broader repository context (such as call graphs and build configurations) requisite for authentic vulnerability reasoning. Training directly upon such fragments introduces severe distribution shifts. To remediate this, we introduce the Repo Context Recovery pipeline. For each annotated snippet, we meticulously preserve the vulnerable determination region while collaboratively reconstructing the peripheral repository architecture utilizing an LLM and static analysis tools. By enforcing strict syntactic validity and successful compilation, we systematically transmute flat snippets into diverse, fully functional projects, conditioning the model for realistic codebase audits.

\paragraph{Feature-Space Deduplication.}
Resolving latent semantic redundancy serves as the final imperative for maximizing the information density of the training corpus and preventing over-fitting to localized conceptual clusters. Recognizing that traditional text-level deduplication (e.g., $n$-gram matching) is structurally insufficient for identifying deep semantic overlap, we adopt an approach predicated on Sparse Autoencoders (SAEs).
Given a corpus $\mathcal{D}=\{x_1,\dots,x_N\}$, we extract hidden-state activations from an intermediate layer and encode them via a pre-trained TopK SAE into sparse feature vectors $\mathbf{z}_i \in \mathbb{R}^{K}$ ($K=65{,}536$). By constructing a neighborhood graph $\mathcal{G}$ where edges indicate cosine similarity exceeding $\tau$, we execute a density-aware pruning algorithm that iteratively excises the node exhibiting the highest local density. The resulting subset is distributed significantly more uniformly within the feature space, explicitly preserving rare topologies while eliminating redundancy.

\subsubsection{Data Quality Control}
\label{sec:data_quality_control}

The integrity of the post-training phase relies heavily on an integrated quality control system designed to collaboratively address individual sample evaluation and systemic multi-task interference.

\paragraph{Data Quality Scoring Model.}
Low-quality instances that exhibit factual inaccuracies or formatting anomalies not only squander computational resources but also demonstrably degrade model alignment~\citep{chen2023alpagasus,liu2024deita}. To systematically counteract this issue, we engineer a Data Quality Scoring Model (DQSM) to evaluate the intrinsic utility of candidate samples. Drawing upon empirical findings that data quality unequivocally supersedes quantity~\citep{liu2024deita}, the DQSM outputs continuous scores across four critical dimensions: (i)Instruction-Response Alignment, which detects off-topic or evasive outputs; (ii)Response Quality, which gauges factual accuracy, depth, and logical coherence; (iii)Instruction Complexity, which quantifies cognitive complexity to prioritize advanced reasoning elicitation; and (iv)Safety Compliance, which evaluates adherence to safety protocols and serves as an indispensable metric for cybersecurity corpora.
Leveraging the LLM-as-a-Judge paradigm, the DQSM computes a composite score:
\begin{align}
    Q(x_i, y_i) = \sum_{k \in \mathcal{K}} \lambda_k \cdot s_i^{(k)}
\end{align}
where dynamic weights $\lambda_k$ can be adapted per task (e.g., elevating $\lambda_{\text{safe}}$ for security data). Unlike rigid binary classifiers, this continuous vector output facilitates highly granular data mixing and dynamic curriculum scheduling.

\paragraph{Mitigating Task Interference in Multi-Task Training.}
Beyond ensuring the intrinsic quality of individual instances, mitigating capability trade-offs across heterogeneous tasks remains a formidable systemic challenge. Improvements in domain-specific tasks frequently induce performance degradation in general capabilities due to gradient conflicts~\citep{borsani2025samgs} and distribution shifts~\citep{blakeney2024sparkjoy}. 
We address this exigency through a synergistic multi-level intervention. At the gradient level, we employ a projection-based gradient surgery approach (PCGrad)~\citep{yu2020pcgrad}, projecting conflicting task gradients onto the normal planes of opposing ones to neutralize destructive interference. At the data level, we implement memory-aware adaptive experience replay~\citep{mssr2026}. By modeling sample retention dynamics grounded in the Ebbinghaus forgetting curve, instances exhibiting rapid degradation receive proportionally higher replay frequencies via prioritized sampling~\citep{schaul2015per}. At the scheduling level, a curriculum learning strategy~\citep{dragomir2025clewr} sequences data chronologically from easy to hard per epoch, naturally fortifying foundational knowledge. Finally, drawing upon Blakeney et al.~\citep{blakeney2024sparkjoy}, we dynamically upsample target-domain instances during the concluding training phases, fortifying specialized proficiency without compromising generalist capabilities.

\subsection{Supervised Fine-Tuning (SFT)}
\label{sec:sft}

With the optimized dataset established, the initial empirical phase commences with Supervised Fine-Tuning. SFT aligns the underlying base model with human instruction typologies and strict formatting conventions by optimizing the standard maximum likelihood estimation (MLE) objective:
\begin{equation}
    \mathcal{L}_{\mathrm{SFT}} = -\mathbb{E}_{(x,y) \sim \mathcal{D}_{\mathrm{SFT}}} \left[ \log \pi_\theta(y \mid x) \right].
\end{equation}
Comprising approximately 300k rigorously filtered instances, the dataset extensively covers cybersecurity analytics and generalized reasoning benchmarks. Crucially, all model responses are explicitly structured to ensure that task-critical variables (e.g., vulnerability identifiers or metric bounds) can be parsed unequivocally. This rigid formatting establishes the essential mechanistic foundation for automated, direct reward computation in the subsequent RL phase.

\subsection{Reinforcement Learning (RL)}
\label{sec:rl}

Having established robust behavioral priors and structural formats via SFT, we transition to Reinforcement Learning (RL) to explicitly elicit advanced analytical reasoning, sophisticated tool manipulation, and overarching policy alignment.

\subsubsection{Multi-Task Reinforcement Learning}

We adopt Group Relative Policy Optimization (GRPO)~\citep{shao2024grpo} as the primary algorithmic engine, prized for its memory efficiency absent a critic model. For each prompt $x$, a cohort of $G$ responses $\{o_i\}_{i=1}^G$ is sampled to yield corresponding rewards $\{r_i\}_{i=1}^G$. The intra-group advantage is dynamically normalized:
\begin{equation}
    \hat{A}_i = \frac{r_i - \mathrm{mean}(\{r_j\}_{j=1}^{G})}{\mathrm{std}(\{r_j\}_{j=1}^{G})}.
\end{equation}
The policy is updated by maximizing the ensuing clipped surrogate objective, regularized by a Kullback-Leibler (KL) divergence penalty:
\begin{equation}
    \mathcal{L}_{\mathrm{GRPO}} = -\mathbb{E} \Biggl[ \frac{1}{G}\sum_{i=1}^{G} \min\!\Bigl(\rho_i\hat{A}_i,\;\mathrm{clip}(\rho_i,1{-}\epsilon,1{+}\epsilon)\hat{A}_i\Bigr) - \lambda\,D_{\mathrm{KL}}(\pi_\theta\|\pi_{\mathrm{ref}}) \Biggr],
\end{equation}
where $\rho_i$ denotes the importance sampling ratio, and $\epsilon$ and $\lambda$ control clipping and penalty magnitudes, respectively. Our framework unifies Reasoning RL, Preference Alignment, and Agentic RL within a shared optimization loop. Across an RL corpus of 120k instances, a difficulty-aware masking mechanism is seamlessly applied to explicitly exclude samples exhibiting intra-group pass rates below 10\% or exceeding 95\%, circumventing degenerate optimization plateaus.

\paragraph{Reasoning RL.}
Inspired by the profound emergent reasoning capabilities documented in DeepSeek-R1~\citep{deepseek2025r1}, we implement RL with Verifiable Rewards (RLVR) to furnish deterministic correctness supervision. For cybersecurity domains, we formalized three objective verification parameters:
\begin{itemize}
    \item \textbf{Root Cause Mapping (RCM)}: Predicting CWE identifiers from CVE semantics, evaluated via a strict binary exact-match indicator: $r_{\mathrm{RCM}} = \mathbf{1}[\hat{y} = y^{*}] \in \{0,1\}$.
    \item \textbf{Vulnerability Severity Prediction (VSP)}: Forecasting aggregate CVSS scores alongside granular metric components: $r_{\mathrm{VSP}} = \gamma\cdot\mathbf{1}[\hat{s}=s^*] + (1-\gamma)\cdot\frac{1}{M}\sum_{m=1}^{M}\mathbf{1}[\hat{v}_m=v_m^*]$.
    \item \textbf{ATT\&CK Technique Extraction (ATE)}: Deriving MITRE techniques from threat intelligence, computed via a set-level F1-score: $r_{\mathrm{ATE}} = \frac{2|\hat{\mathcal{T}}\cap\mathcal{T}^*|}{|\hat{\mathcal{T}}|+|\mathcal{T}^*|}$. Accommodating partial credit, this continuous metric precipitates demonstrably smoother gradient flows than rigid binary variants.
\end{itemize}
Advancing beyond terminal outcomes, we actively inject process-level reward paradigms into the RLVR framework. Evaluating intermediate reasoning steps provides dense localized signals, exponentially accelerating policy convergence.

\paragraph{Agentic RL.}
While RLVR dramatically escalates single-turn analytical proficiency, modern cybersecurity workflows fundamentally demand protracted interactive dexterity, multi-step planning, and diverse tool invocation. Agentic RL directly addresses this gap by optimizing policies in tool-augmented operational environments, evaluated strictly upon end-to-end task completion. Scenarios encompass automated CTF solving, red/blue-team simulation exercises, and active vulnerability remediation. 
To bypass the lack of native systemic integration for highly specialized operational tooling (e.g., fuzzers, debuggers), we are progressively networking the underlying architecture into these specialized ecosystems. Because agentic rewards are intrinsically sparse and temporally delayed, this framework seamlessly complements the localized, high-density reward topologies inherent to RLVR, facilitating a joint optimization of analytical precision and macroscopic planning.

\paragraph{ActionRL.}
To complement GRPO, we incorporate ActionRL~\citep{qoder2026actionrl}, a divergence-aware optimization algorithm designed for fine-grained preference alignment. Conventional sequence-level objectives such as DPO~\citep{dpo23} distribute the contrastive loss uniformly across all tokens, thereby penalizing the shared semantic prefix between chosen and rejected responses—a region that carries no discriminative signal. This indiscriminate credit assignment is particularly detrimental in security-critical code synthesis, where exploitable flaws often originate from a single localized decision point rather than from the response as a whole. ActionRL mitigates this issue by explicitly identifying the Branch Divergence Point defined as the first token at which the chosen and rejected trajectories bifurcate, and confining gradient attribution strictly to the divergent suffix. By restricting supervision to the truly contrastive region, ActionRL eliminates spurious penalties on legitimate shared context and alleviates the over-conservatism commonly observed in full-sequence preference optimization, yielding sharper and more targeted alignment signals.

\subsubsection{Self-Evolution and Self-Play}
\label{sec:self_evo}

While RLVR and Agentic RL provide exceptionally dense supervisory signals, their capability ceilings are ultimately bounded by the predefined scope of fixed reward templates and static environments. To transcend these human-engineered limitations and reinforce zero-shot robustness, the terminal mechanism for reinforcing model alignment relies upon sustained, autonomous self-improvement.

\paragraph{Self-Evolving via Adversarial RL with Code-Grounded Verification.}
To facilitate autonomous continuous evolution tailored for cybersecurity, we implement an adversarial loop predicated on SEAS~\citep{diao2025seas}. This paradigm orchestrates an iterative engagement between a Red Team model (attacker) and a Target model (defender). Adapting the co-evolutionary framework from Code-A1~\citep{wang2026codea1} alongside GRPO for memory-efficient updating, the opposing entities are instantiated as independent architectures burdened by diametrically opposing objectives. The Red Team secures positive rewards solely upon successfully executing a sandbox vulnerability exploit, whereas the Target receives analogous rewards only when formulating functional security patches that definitively obstruct the attack. This dual-model dichotomy fundamentally circumvents the systemic self-collusion vulnerabilities intrinsic to solitary self-play frameworks. Synergized with a Mistake Book mechanism for experience replay mapping, this generates an inviolable self-evolutionary loop defined by data synthesis, code-grounded verification, and unremitting adversarial capability escalation.

\paragraph{Vulnerability Self-Play.}
Authentic, security-critical vulnerabilities seldom manifest as rudimentary syntax errors; rather, they predominantly emerge when adversaries adroitly circumvent entrenched safeguards, purposefully engineering configurations to subvert orthodox detection heuristics. This establishes an explicit game-theoretic duality interlinking vulnerability integration and detection. We encapsulate this adversarial dynamic via a self-play framework orchestrating a vulnerability writer against a corresponding vulnerability detector. The writer iteratively formulates intricately constructed exploits designed specifically to target the analytical blind spots of contemporary heuristics. Concurrently, the detector undergoes sustained conditioning aimed at identifying and analytically deconstructing these permutations. As both entities progressively escalate in procedural sophistication, they iteratively synthesize highly formidable supervisory signals. By natively embedding this self-play loop across both data construction parameters and advanced RL pipelines, we guarantee an exponentially expanding curriculum strictly defined by high-fidelity security permutations.

\section{Evaluation} \label{sec:experiments}

A domain-specialized model must excel not only on security-specific tasks, but
also retain the broad general capabilities that make it practically useful.
Evaluating on a single axis risks overfitting to narrow task distributions and
masking regressions elsewhere. We therefore adopt a multi-dimensional
evaluation protocol spanning general-purpose benchmarks across languages,
reasoning modalities, and math tasks, alongside a comprehensive suite of
cybersecurity-specific benchmarks. This design ensures that gains in domain
performance are not achieved at the expense of general intelligence, and that
the model exhibits genuine compositional competence rather than superficial
pattern matching on a handful of familiar benchmarks.

\paragraph{Training Setup.}
XekRung-8B is trained through a three-stage pipeline built upon the
Qwen3-8B base model. Continued pre-training (CPT) first exposes the base
model to a curated cybersecurity corpus, injecting domain-specific knowledge
and terminology. Supervised fine-tuning (SFT) is then applied to the
resulting CPT checkpoint using full-parameter tuning on a mixed corpus of
cybersecurity-specific and general-purpose instruction data with a causal
language modeling objective and cosine-decay learning rate schedule.
Finally, GRPO-based multi-task reinforcement learning further optimizes the
SFT checkpoint; the RL corpus spans cybersecurity-specific tasks with
verifiable rewards and general-domain tasks for capability retention.

\paragraph{Baselines.}
We compare against seven baseline models organized by model family:
same-scale and larger Qwen-series models (Qwen3-8B, Qwen3-14B, Qwen3.5-9B),
Llama-series models (Llama-3.1-8B-Instruct, Llama-3.3-70B-Instruct,
Llama-Primus-Reasoning-8B), and existing cybersecurity-specialized models
(Foundation-Sec-8B-Reasoning, SecGPT-14B). This selection enables us to
assess same-scale gains, cross-scale competitiveness, and domain-specific
effectiveness simultaneously.

For all evaluations, we use greedy decoding (temperature\,=\,0) to ensure
deterministic and reproducible results. General-purpose benchmarks are
evaluated using the lm-evaluation-harness framework with the number of
in-context examples specified in Table~\ref{tab:general-benchmarks};
cybersecurity benchmarks follow the same protocol with per-benchmark
configurations detailed in Table~\ref{tab:security-benchmarks}.


\begin{table*}[t]
\centering
\caption{General-purpose benchmark suite. ``Shots'' indicates the number of
in-context examples used during evaluation.}
\label{tab:general-benchmarks}
\small
\begin{minipage}[t]{0.48\textwidth}
\centering
\resizebox{\linewidth}{!}{%
\begin{tabular}{llcc}
\toprule
\textbf{Category} & \textbf{Benchmark} & \textbf{Shots} & \textbf{Metric} \\
\midrule
\multirow{6}{*}{\shortstack[l]{English\\Knowledge}}
 & MMLU~\citep{hendrycks2021mmlu} & 5 & Acc \\
 & MMLU-Pro~\citep{wang2024mmlupro} & 5 & Acc \\
 & MMLU-Redux~\citep{garg2024mmluredux} & 5 & Acc \\
 & ARC-Challenge~\citep{clark2018arc} & 25 & Acc \\
 & GPQA~\citep{rein2024gpqa} & 5 & Acc \\
 & Xiezhi-EN~\citep{gu2023xiezhi} & 3 & Acc \\
\midrule
\multirow{4}{*}{\shortstack[l]{English\\Reading}}
 & TriviaQA~\citep{joshi2017triviaqa} & 5 & EM \\
 & NarrativeQA~\citep{kocisky2018narrativeqa} & 3 & F1 \\
 & OpenBookQA~\citep{mihaylov2018openbookqa} & 5 & Acc \\
 & SQuAD 2.0~\citep{rajpurkar2018squad} & 3 & F1 \\
\midrule
\multirow{3}{*}{\shortstack[l]{English\\Language}}
 & LAMBADA~\citep{paperno2016lambada} & 0 & Acc \\
 & StoryCloze~\citep{mostafazadeh2017storycloze} & 0 & Acc \\
 & BoolQ~\citep{clark2019boolq} & 5 & Acc \\
\midrule
\multirow{7}{*}{\shortstack[l]{English\\Reasoning}}
 & BBH~\citep{suzgun2023bbh} & 3 & EM \\
 & HellaSwag~\citep{zellers2019hellaswag} & 10 & Acc \\
 & WinoGrande~\citep{sakaguchi2021winogrande} & 5 & Acc \\
 & PIQA~\citep{bisk2020piqa} & 5 & Acc \\
 & SocialIQA~\citep{sap2019socialiqa} & 5 & Acc \\
 & CommonsenseQA~\citep{talmor2019commonsenseqa} & 7 & Acc \\
\bottomrule
\end{tabular}}
\end{minipage}\hfill
\begin{minipage}[t]{0.48\textwidth}
\centering
\resizebox{\linewidth}{!}{%
\begin{tabular}{llcc}
\toprule
\textbf{Category} & \textbf{Benchmark} & \textbf{Shots} & \textbf{Metric} \\
\midrule
\multirow{3}{*}{\shortstack[l]{Chinese\\Language}}
 & CSEM~\citep{cui2022csem} & 5 & Acc \\
 & CHID~\citep{zheng2020chid} & 5 & Acc \\
 & WPLC~\citep{ge2021wplc} & 0 & Acc \\
\midrule
\multirow{3}{*}{\shortstack[l]{Chinese\\Knowledge}}
 & CMMLU~\citep{li2023cmmlu} & 5 & Acc \\
 & C-Eval~\citep{huang2023ceval} & 5 & Acc \\
 & Xiezhi-CN~\citep{gu2023xiezhi} & 3 & Acc \\
\midrule
\multirow{3}{*}{\shortstack[l]{Chinese\\Reasoning}}
 & C$^3$~\citep{sun2020c3} & 3 & Acc \\
 & CLUEWSC~\citep{xu2020clue} & 5 & Acc \\
 & OCNLI~\citep{hu2020ocnli} & 5 & Acc \\
\midrule
\multirow{4}{*}{\shortstack[l]{English\\Math}}
 & MATH~\citep{hendrycks2021math} & 4 & Acc \\
 & TAL-SCQ5K-EN~\citep{zheng2021tal} & 3 & Acc \\
 & DROP~\citep{dua2019drop} & 3 & F1 \\
 & TheoremQA~\citep{chen2023theoremqa} & 5 & Acc \\
\midrule
\multirow{3}{*}{\shortstack[l]{Chinese\\Math}}
 & TAL-SCQ5K-CN~\citep{zheng2021tal} & 3 & Acc \\
 & GSM8K-ZH~\citep{cobbe2021gsm8k} & 5 & Acc \\
 & Ape210K~\citep{zhao2020ape210k} & 5 & Acc \\
\bottomrule
\end{tabular}}
\end{minipage}

\end{table*}

\subsection{General-Purpose Benchmarks}
\label{sec:eval-general}

We evaluate on 35 general-purpose benchmarks organized into 9 categories, as
listed in Table~\ref{tab:general-benchmarks}. The suite spans English and
Chinese knowledge, reading and language understanding, commonsense and formal
reasoning, and mathematics, providing a comprehensive profile of the model's
general capabilities.

Table~\ref{tab:general-results} presents results across all 35 general-purpose benchmarks.

\begin{table*}[!t]
\centering
\caption{General-purpose benchmark results across all 35 benchmarks (\%). Best results among $\leq$14B models are in \textbf{bold} and second-best are \underline{underlined}. A dash (–) indicates that the model failed to produce valid outputs due to output-format incompatibility, and such entries are excluded from comparison.}
\label{tab:general-results}
\setlength{\tabcolsep}{2.0pt}
\renewcommand{\arraystretch}{0.92}
\scriptsize
\renewcommand\theadalign{cc}
\renewcommand\theadfont{\bfseries}
\renewcommand\theadgape{\Gape[2pt]}

\resizebox{\textwidth}{!}{%
\begin{tabular}{@{} l l *{9}{c} @{}}
\toprule
\thead[l]{Category} & \thead[l]{Benchmark}
 & \thead{Qwen3-8B}
 & \thead{Qwen3-14B}
 & \thead{Qwen3.5-9B}
 & \thead{Llama-3.1-\\8B-Instruct}
 & \thead{Llama-3.3-\\70B-Instruct}
 & \thead{Llama-Primus-\\Reasoning-8B}
 & \thead{Foundation-Sec-\\8B-Reasoning}
 & \thead{SecGPT-\\14B}
 & \thead{XekRung-\\8B} \\
\midrule
\multirow{6}{*}{EN Knowledge}
 & MMLU & 78.23 & \underline{81.94} & \textbf{84.58} & 69.40 & 85.86 & 45.55 & 68.30 & 73.73 & 78.58 \\
 & MMLU-Pro & 62.20 & 68.60 & \textbf{75.25} & 37.55 & 67.90 & - & 38.80 & 55.35 & \underline{70.40} \\
 & MMLU-Redux & 77.53 & \underline{80.47} & \textbf{83.43} & - & 81.73 & 21.80 & - & 69.50 & 78.77 \\
 & ARC-Challenge & 93.09 & \underline{94.54} & \textbf{94.80} & 79.30 & 95.99 & - & - & 90.44 & 93.43 \\
 & GPQA & 41.92 & \textbf{48.48} & \underline{45.96} & 30.40 & 51.52 & - & 31.70 & 13.13 & 43.43 \\
 & Xiezhi-EN & \underline{68.55} & \textbf{70.85} & 68.50 & 21.70 & 68.05 & 35.75 & - & 67.10 & 68.45 \\
\midrule
\multirow{4}{*}{\shortstack[l]{English\\Reading}}
 & TriviaQA & 57.90 & \textbf{66.50} & 59.45 & \underline{62.40} & 83.70 & 62.15 & 52.75 & 42.40 & 54.70 \\
 & NarrativeQA & 29.85 & \textbf{35.75} & 30.10 & 33.20 & 53.35 & 26.55 & 25.70 & \underline{35.35} & 27.90 \\
 & OpenBookQA & 91.00 & \underline{94.00} & \textbf{94.60} & 26.40 & 96.00 & 30.80 & - & 89.80 & 93.80 \\
 & SQuAD 2.0 & 69.60 & \underline{74.40} & \textbf{79.25} & 29.10 & 71.85 & 38.55 & 46.30 & 71.80 & 56.00 \\
\midrule
\multirow{3}{*}{\shortstack[l]{English\\Language}}
 & LAMBADA & 44.20 & \textbf{56.70} & \underline{45.10} & - & 62.30 & 19.85 & 11.15 & 41.75 & 32.15 \\
 & StoryCloze & 97.09 & \textbf{98.81} & 97.02 & - & 99.47 & 38.45 & - & \underline{97.55} & 97.42 \\
 & BoolQ & 86.24 & 86.70 & 85.66 & 80.00 & 89.42 & 45.29 & - & \textbf{87.95} & \underline{87.77} \\
\midrule
\multirow{6}{*}{\shortstack[l]{English\\Reasoning}}
 & BBH & 69.95 & 69.05 & \textbf{77.45} & 67.40 & 77.40 & 37.05 & 69.90 & 64.75 & \underline{72.40} \\
 & HellaSwag & 81.33 & \underline{86.70} & \textbf{88.51} & 80.90 & 89.80 & 31.28 & - & 82.69 & 77.43 \\
 & WinoGrande & 68.75 & \underline{76.16} & \textbf{82.64} & - & 81.06 & 40.41 & - & 72.45 & 73.24 \\
 & PIQA & 87.11 & \textbf{89.88} & \underline{88.41} & 80.30 & 91.89 & 46.57 & - & 87.21 & 86.40 \\
 & SocialIQA & 74.51 & \textbf{77.43} & \underline{76.56} & 18.01 & 80.86 & 27.02 & - & 73.08 & 73.39 \\
 & CommonsenseQA & \underline{82.56} & 82.31 & \textbf{82.64} & - & 85.26 & 40.46 & - & 79.52 & 80.59 \\
\midrule
\multirow{3}{*}{\shortstack[l]{Chinese\\Language}}
 & CSEM & 90.34 & \textbf{93.14} & \underline{92.71} & 43.81 & 90.25 & 52.03 & 14.41 & 91.02 & 87.71 \\
 & CHID & 80.27 & \underline{84.92} & \textbf{85.61} & - & 82.37 & - & 25.42 & 78.57 & 77.62 \\
 & WPLC & 19.95 & \textbf{24.50} & 10.95 & 5.00 & 23.20 & 4.25 & 4.70 & \underline{20.15} & 12.50 \\
\midrule
\multirow{3}{*}{\shortstack[l]{Chinese\\Knowledge}}
 & CMMLU & 77.45 & \underline{82.10} & \textbf{84.35} & 34.45 & 72.25 & 43.15 & - & 72.35 & 77.05 \\
 & C-Eval & 78.59 & \underline{81.78} & \textbf{82.16} & 31.30 & 69.96 & 36.80 & 11.60 & 67.51 & 76.65 \\
 & Xiezhi-CN & 78.15 & \underline{79.60} & \textbf{80.20} & 31.30 & 76.45 & 55.35 & - & 77.40 & 77.60 \\
\midrule
\multirow{3}{*}{\shortstack[l]{Chinese\\Reasoning}}
 & C$^3$ & 92.00 & \underline{94.74} & \textbf{95.51} & 28.38 & 96.88 & - & 25.48 & 93.21 & 91.18 \\
 & CLUEWSC & 90.37 & \textbf{92.73} & \underline{92.01} & 77.97 & 90.68 & 29.82 & 18.24 & 88.73 & 87.70 \\
 & OCNLI & 68.81 & 69.72 & \textbf{74.40} & 43.81 & 71.59 & 30.60 & 18.33 & 67.66 & \underline{69.96} \\
\midrule
\multirow{4}{*}{\shortstack[l]{English\\Math}}
 & MATH & 71.05 & \textbf{74.70} & 66.90 & 51.90 & 69.55 & 35.30 & 43.30 & 44.70 & \underline{73.50} \\
 & TAL-SCQ5K-EN & 89.00 & \textbf{90.95} & \underline{90.10} & - & 86.40 & 43.85 & - & 75.95 & 88.75 \\
 & DROP & 79.90 & \underline{89.95} & \textbf{91.25} & 70.20 & 92.10 & 57.70 & 61.20 & 88.35 & 78.10 \\
 & TheoremQA & 40.12 & \underline{42.88} & \textbf{49.25} & 26.00 & 38.88 & 5.37 & 22.25 & 23.50 & 41.63 \\
\midrule
\multirow{3}{*}{\shortstack[l]{Chinese\\Math}}
 & TAL-SCQ5K-CN & 75.50 & \underline{79.75} & \textbf{84.50} & - & 65.65 & - & - & 45.55 & 77.70 \\
 & GSM8K-ZH & 87.87 & \textbf{91.28} & \underline{91.21} & 70.51 & 92.57 & 68.76 & 59.97 & 89.39 & 89.76 \\
 & Ape210K & 85.00 & \underline{87.55} & \textbf{90.10} & 49.35 & 79.15 & 53.30 & 47.05 & 81.00 & 85.20 \\
\midrule
\multicolumn{2}{l}{\textbf{Overall Avg}}
 & 73.31 & \underline{77.13} & \textbf{77.17} & - & 77.47 & - & - & 68.59 & 72.54 \\
\bottomrule
\end{tabular}}
\end{table*}

\subsection{Cybersecurity Benchmarks}
\label{sec:eval-security}

We evaluate on 15 cybersecurity benchmarks spanning cyber threat intelligence,
vulnerability analysis, broad security knowledge, and applied security
engineering, as listed in Table~\ref{tab:security-benchmarks}. Several
agentic and code-level benchmarks are under active integration and will be
included in subsequent evaluations. We note that the results presented here reflect evaluations at the current 8B model scale; comprehensive results on agentic benchmarks—where long-horizon planning and tool orchestration impose substantially higher demands on model capacity—will be reported alongside our forthcoming larger-scale model. Table~\ref{tab:security-results} presents the results across all 15 cybersecurity benchmarks listed above.

\begin{table}[t]
\centering
\caption{Cybersecurity benchmark suite.}
\label{tab:security-benchmarks}
\scriptsize

\renewcommand{\arraystretch}{1.25}

\begin{tabular}{@{} l l >{\raggedright\arraybackslash}m{7.2cm} @{}}
\toprule
\textbf{Benchmark} & \textbf{Task Type} & \textbf{Description} \\
\midrule
\addlinespace[3pt]
CSE-Benchmark~\citep{wang2025csebenchmark}
 & Multiple choice & Expert-level evaluation across 18 competency domains \\
CTIBench-MCQ~\citep{alam2024ctibench}
 & Multiple choice & Cyber threat intelligence knowledge \\
CTIBench-RCM~\citep{alam2024ctibench}
 & CVE$\to$CWE mapping & Vulnerability root cause mapping \\
CTIBench-VSP~\citep{alam2024ctibench}
 & Severity prediction & CVSS vector string prediction \\
CTIBench-ATE~\citep{alam2024ctibench}
 & Technique extraction & MITRE ATT\&CK TTP extraction from threat reports \\
CyberMetric-10kQ~\citep{tihanyi2024cybermetric}
 & Multiple choice & Security standards and publications knowledge (10k Qs) \\
CyberMetric-2kQ~\citep{tihanyi2024cybermetric}
 & Multiple choice & Security standards and publications knowledge (2k Qs) \\
CyberMetric-500Q~\citep{tihanyi2024cybermetric}
 & Multiple choice & Security standards and publications knowledge (500 Qs) \\
CyberMetric-80Q~\citep{tihanyi2024cybermetric}
 & Multiple choice & Security standards and publications knowledge (80 Qs) \\
CISSP-MC-ZH
 & Multiple choice & CISSP certification questions (Chinese) \\
CISSP-SC-EN
 & Multiple choice & CISSP certification questions (English) \\
CISSP-SC-ZH
 & Multiple choice & CISSP certification questions (Chinese) \\
MMLU-Computer-Security~\citep{hendrycks2021mmlu}
 & Multiple choice & Computer security subset of MMLU \\
SecBench-MCQ~\citep{liu2024secbench}
 & Multiple choice & Security knowledge evaluation (multiple choice) \\
SecEval~\citep{li2023seceval}
 & Multiple choice & 9-domain security knowledge evaluation \\
\bottomrule
\end{tabular}

\end{table}

\begin{table*}[t]
\centering
\caption{Cybersecurity benchmark results (\%). Best results among $\leq$14B models are in \textbf{bold} and second-best are \underline{underlined}. All results are evaluated under our unified evaluation framework.}
\label{tab:security-results}
\setlength{\tabcolsep}{2.0pt}
\renewcommand{\arraystretch}{1.1}
\scriptsize

\renewcommand\theadalign{cc}       
\renewcommand\theadfont{\bfseries}
\renewcommand\theadgape{\Gape[2pt]} 
\renewcommand\cellgape{\Gape[2pt]}

\resizebox{\textwidth}{!}{%
\begin{tabular}{@{} l l *{9}{c} @{}}
\toprule
\thead{Category} & \thead{Benchmark}
 & \thead{Qwen3-8B}
 & \thead{Qwen3-14B}
 & \thead{Qwen3.5-9B}
 & \thead{Llama-3.1\\-8B-Instruct}
 & \thead{Llama-3.3\\-70B-Instruct}
 & \thead{Llama-Primus\\-Reasoning-8B}
 & \thead{Foundation-Sec\\-8B-Reasoning}
 & \thead{SecGPT-14B}
 & \thead{XekRung-8B} \\
\midrule
\multirow{5}{*}{Threat Intel}
 & CSE-Benchmark & 80.17 & \textbf{83.06} & 80.10 & 66.80 & 82.42 & 55.44 & 64.06 & 80.33 & \underline{80.77} \\
 & CTIBench-MCQ & 37.64 & 64.60 & \underline{67.08} & 52.00 & 64.24 & 65.16 & 50.80 & 57.64 & \textbf{70.52} \\
 & CTIBench-RCM & 54.50 & 61.20 & 63.40 & 57.90 & 65.70 & 73.70 & 74.10 & \underline{77.10} & \textbf{77.30} \\
 & CTIBench-VSP & 84.15 & 84.65 & 75.55 & 81.73 & 85.17 & 80.46 & \underline{86.59} & 3.71 & \textbf{92.12} \\
 & CTIBench-ATE & 25.41 & 40.88 & 50.49 & 17.02 & 47.46 & 27.22 & \underline{54.45} & 14.80 & \textbf{73.39} \\
\midrule
\multirow{3}{*}{Vuln Analysis}
 & MMLU-CompSec & 81.00 & \textbf{84.00} & 79.00 & 50.00 & 78.00 & 56.00 & 67.00 & 72.00 & \underline{83.00} \\
 & SecEval & 62.14 & 66.68 & 68.74 & 38.63 & 73.79 & 44.68 & 59.26 & \underline{69.25} & \textbf{71.04} \\
 & SecBench-MCQ & 88.57 & \textbf{90.59} & \underline{89.67} & 72.93 & 86.89 & 70.59 & 75.53 & 85.64 & 88.86 \\
\midrule
\multirow{4}{*}{Sec Knowledge}
 & CyberMetric-10kQ & 84.87 & 86.31 & \textbf{86.47} & 80.50 & 87.67 & 81.99 & 81.10 & 84.12 & \underline{86.32} \\
 & CyberMetric-2kQ & 88.70 & 89.80 & \textbf{91.25} & 84.25 & 91.90 & 85.55 & 84.15 & 88.65 & \underline{90.70} \\
 & CyberMetric-500Q & 88.80 & 91.60 & \underline{91.80} & 83.00 & 94.20 & 85.80 & 86.20 & 90.60 & \textbf{92.80} \\
 & CyberMetric-80Q & 93.75 & \textbf{96.25} & \textbf{96.25} & 88.75 & 96.25 & 91.25 & 93.75 & \underline{95.00} & \underline{95.00} \\
\midrule
\multirow{3}{*}{Certification}
 & CISSP-MC-ZH & 55.81 & \underline{58.29} & \textbf{60.99} & 38.13 & 53.03 & 21.04 & 23.30 & 49.96 & 57.71 \\
 & CISSP-SC-EN & 75.93 & \textbf{79.66} & \underline{78.63} & 62.38 & 79.53 & 59.34 & 66.71 & 75.96 & 77.11 \\
 & CISSP-SC-ZH & 78.66 & \textbf{82.25} & \underline{79.89} & 54.95 & 77.83 & 44.99 & 61.08 & 78.36 & 79.01 \\
\midrule
\multicolumn{2}{l}{\textbf{Overall Avg}}
 & 72.01 & \underline{77.32} & 77.29 & 61.93 & 77.61 & 62.88 & 68.54 & 68.21 & \textbf{81.04} \\
\bottomrule
\end{tabular}
}
\end{table*}


\subsection{Analysis}
\label{sec:eval-analysis}

\paragraph{Same-Scale Superiority.}
At the 8B parameter scale, XekRung-8B outperforms both general-purpose
baselines (Qwen3-8B, Llama-3.1-8B-Instruct) and security-specialized
baselines (Foundation-Sec-8B-Reasoning, Llama-Primus-Reasoning-8B)
across both benchmark suites, demonstrating that our complete
three-stage training paradigm---CPT, multi-task SFT, and GRPO-based
RL---yields substantial gains over partial-pipeline approaches and over
models trained exclusively on security data without preserving general
capabilities.

\paragraph{Cross-Scale Competitiveness.}
Beyond same-scale comparisons, XekRung-8B achieves competitive or superior
performance relative to Qwen3-14B (nearly $2\times$ the parameter count)
across several benchmark categories, and approaches the performance of
Llama-3.3-70B-Instruct on multiple security-specific benchmarks.
Notably, XekRung-8B also outperforms the larger SecGPT-14B on security
benchmarks by a significant margin ($81.04$ vs.~$68.21$ overall),
further highlighting the high cost-effectiveness of domain compression:
a well-trained 8B model can match or exceed models with several times more
parameters on domain-specific tasks, while offering substantially faster
inference---a critical advantage for latency-sensitive security operations
such as real-time alert triage and interactive threat analysis.

\paragraph{Training Progression.}
Each training stage contributes meaningfully to the final model. CPT injects
domain-specific knowledge by exposing the model to a curated cybersecurity
corpus, establishing foundational understanding of security concepts,
terminology, and ontology. Multi-task SFT then builds on this foundation to
develop instruction-following and task-specific formatting, enabling the model
to respond effectively to diverse query types. Finally, GRPO-based multi-task
RL sharpens analytical reasoning through verifiable-reward reinforcement
learning and extends agentic capabilities via AgenticRL, allowing the
model to perform complex multi-step security analysis tasks. This progressive
training paradigm ensures that the model achieves deep domain expertise while
maintaining broad general-purpose competence.

\paragraph{Scaling Implications.}
XekRung-8B's strong performance at a relatively small parameter scale, combined
with the systematic improvements at each training stage, provides confidence
that our training methodology will transfer effectively to larger model
architectures. These results lay the groundwork for extending domain-specialized
capabilities to larger backbones, with the potential to further advance both
general reasoning and security-specific expertise at scale.

\section{Conclusion}\label{sec:conclusion}
In this technical report, we introduce XekRung,  a frontier large language model  tailored for the cybersecurity.
We inject cybersecurity knowledge, understanding and reasoning ability into the model via continued pre-training, supervised fine-tuning and  reinforcement learning  respectively.
By adopting a complete three-stage training paradigm based on the Qwen family, XekRung demonstrates outstanding performance in a wide range of cybersecurity tasks, while maintaining strong performance on general benchmarks.
In the future, we will further expand the model’s agentic capabilities and investigate the potential of self-evolution for cybersecurity.
We also plan to pursue broader cybersecurity research and applications, contributing to the continued advancement of AI for Security.

\section{Authors}
\label{sec:author_and_ack}

\paragraph{\textbf{Core Contributors:}}
Jiutian Zeng$^*$, Junjie Li$^*$, Chengwei Dai$^*$, Jie Liang$^*$, Zhaoyu Hu$^*$, Yiliang Zhang, Ziang Weng, Longtao Huang

{\scriptsize ($^*$ denotes equal contribution)}

\paragraph{\textbf{Contributors:}}
Dongjie Zhang, Libin Dong, Yang Ge, Yuanda Wang, Kaiwen Lv Kacuila, Bingyu Zhu, Jing Wang, Jin Xu

\clearpage
\bibliography{biblio}

\begin{thebibliography}{87}
\providecommand{\natexlab}[1]{#1}
\providecommand{\url}[1]{\texttt{#1}}
\expandafter\ifx\csname urlstyle\endcsname\relax
  \providecommand{\doi}[1]{doi: #1}\else
  \providecommand{\doi}{doi: \begingroup \urlstyle{rm}\Url}\fi

\bibitem[Agarwal et~al.(2024)Agarwal, Vieillard, Zhou, Stanczyk, Ramos~Garea, Geist, and Bachem]{agarwal2024policy}
Rishabh Agarwal, Nino Vieillard, Yongchao Zhou, Piotr Stanczyk, Sabela Ramos~Garea, Matthieu Geist, and Olivier Bachem.
\newblock On-policy distillation of language models: Learning from self-generated mistakes.
\newblock In \emph{The Twelfth International Conference on Learning Representations}, 2024.

\bibitem[Alam et~al.(2024)Alam, Bhusal, Nguyen, and Rastogi]{alam2024ctibench}
Md~Tanvirul Alam, Dipkamal Bhusal, Le~Nguyen, and Nidhi Rastogi.
\newblock {CTIBench}: A benchmark for evaluating {LLMs} in cyber threat intelligence.
\newblock In \emph{The Thirty-eighth Conference on Neural Information Processing Systems, Datasets and Benchmarks Track}, 2024.

\bibitem[{Anthropic}(2026)]{anthropic2026mythos}
{Anthropic}.
\newblock Assessing {Claude Mythos Preview}'s cybersecurity capabilities, 2026.
\newblock URL \url{https://red.anthropic.com/2026/mythos-preview/}.

\bibitem[Bai et~al.(2023)]{bai2023qwen}
Jinze Bai et~al.
\newblock Qwen technical report.
\newblock \emph{arXiv preprint arXiv:2309.16609}, 2023.
\newblock URL \url{https://arxiv.org/abs/2309.16609}.

\bibitem[Bisk et~al.(2020)Bisk, Zellers, Le~Bras, Gao, and Choi]{bisk2020piqa}
Yonatan Bisk, Rowan Zellers, Ronan Le~Bras, Jianfeng Gao, and Yejin Choi.
\newblock {PIQA}: Reasoning about physical commonsense in natural language.
\newblock In \emph{Proceedings of the AAAI Conference on Artificial Intelligence}, volume~34, pp.\  7432--7439. AAAI Press, 2020.
\newblock \doi{10.1609/aaai.v34i05.6239}.
\newblock URL \url{https://doi.org/10.1609/aaai.v34i05.6239}.

\bibitem[Blakeney et~al.(2024)Blakeney, Paul, Larsen, Owen, and Frankle]{blakeney2024sparkjoy}
Cody Blakeney, Mansheej Paul, Brett~W. Larsen, Sean Owen, and Jonathan Frankle.
\newblock Does your data spark joy? performance gains from domain upsampling at the end of training.
\newblock \emph{CoRR}, abs/2406.03476, 2024.
\newblock \doi{10.48550/ARXIV.2406.03476}.
\newblock URL \url{https://doi.org/10.48550/arXiv.2406.03476}.

\bibitem[Borsani et~al.(2025)Borsani, Rosani, Nicosia, and Di~Fatta]{borsani2025samgs}
Thomas Borsani, Andrea Rosani, Giuseppe Nicosia, and Giuseppe Di~Fatta.
\newblock Gradient similarity surgery in multi-task deep learning.
\newblock In \emph{Proceedings of the European Conference on Machine Learning and Principles and Practice of Knowledge Discovery in Databases (ECMLPKDD)}, 2025.
\newblock arXiv:2506.06130.

\bibitem[Chen et~al.(2023{\natexlab{a}})Chen, Li, Yan, Wang, Gunaratna, Yadav, Tang, Srinivasan, Zhou, Huang, and Jin]{chen2023alpagasus}
Lichang Chen, Shiyang Li, Jun Yan, Hai Wang, Kalpa Gunaratna, Vikas Yadav, Zheng Tang, Vijay Srinivasan, Tianyi Zhou, Heng Huang, and Hongxia Jin.
\newblock Alpagasus: Training a better alpaca with fewer data.
\newblock \emph{arXiv preprint arXiv:2307.08701}, 2023{\natexlab{a}}.

\bibitem[Chen et~al.(2023{\natexlab{b}})Chen, Yin, Ku, Lu, Wan, Ma, Xu, Wang, and Xia]{chen2023theoremqa}
Wenhu Chen, Ming Yin, Max Ku, Pan Lu, Yixin Wan, Xueguang Ma, Jianyu Xu, Xinyi Wang, and Tony Xia.
\newblock {T}heorem{QA}: A theorem-driven question answering dataset.
\newblock In \emph{Proceedings of the 2023 Conference on Empirical Methods in Natural Language Processing}, pp.\  7889--7901, Singapore, December 2023{\natexlab{b}}. Association for Computational Linguistics.
\newblock \doi{10.18653/v1/2023.emnlp-main.489}.
\newblock URL \url{https://aclanthology.org/2023.emnlp-main.489/}.

\bibitem[Clark et~al.(2019)Clark, Lee, Chang, Kwiatkowski, Collins, and Toutanova]{clark2019boolq}
Christopher Clark, Kenton Lee, Ming-Wei Chang, Tom Kwiatkowski, Michael Collins, and Kristina Toutanova.
\newblock Boolq: Exploring the surprising difficulty of natural yes/no questions.
\newblock \emph{Proceedings of NAACL}, 2019.

\bibitem[Clark et~al.(2018)Clark, Cowhey, Etzioni, Khot, Sabharwal, Schoenick, and Tafjord]{clark2018arc}
Peter Clark, Isaac Cowhey, Oren Etzioni, Tushar Khot, Ashish Sabharwal, Carissa Schoenick, and Oyvind Tafjord.
\newblock Think you have solved question answering? try arc, the ai2 reasoning challenge.
\newblock \emph{arXiv preprint arXiv:1803.05457}, 2018.

\bibitem[{Clouditera}(2025)]{secgpt}
{Clouditera}.
\newblock Secgpt: The world's first open-source large language model for cybersecurity.
\newblock \url{https://huggingface.co/clouditera/secgpt}, 2025.
\newblock Accessed: 2025-04.

\bibitem[Cobbe et~al.(2021)Cobbe, Kosaraju, Bavarian, Chen, Jun, Kaiser, Plappert, Tworek, Hilton, Nakano, Hesse, and Schulman]{cobbe2021gsm8k}
Karl Cobbe, Vineet Kosaraju, Mohammad Bavarian, Mark Chen, Heewoo Jun, Lukasz Kaiser, Matthias Plappert, Jerry Tworek, Jacob Hilton, Reiichiro Nakano, Christopher Hesse, and John Schulman.
\newblock Training verifiers to solve math word problems.
\newblock \emph{arXiv preprint arXiv:2110.14168}, 2021.
\newblock \doi{10.48550/arXiv.2110.14168}.
\newblock URL \url{https://arxiv.org/abs/2110.14168}.

\bibitem[{CyberNative AI LLC}(2024)]{colibria2025colibri8b}
{CyberNative AI LLC}.
\newblock Colibri\_8b\_v0.1: Conversational cybersecurity model.
\newblock Hugging Face model repository, May 2024.
\newblock URL \url{https://huggingface.co/CyberNative-AI/Colibri_8b_v0.1}.
\newblock Llama-3-8B cybersecurity dialogue fine-tune; published 15 May 2024.

\bibitem[{DeepSeek-AI}(2024)]{deepseek2024v3}
{DeepSeek-AI}.
\newblock {DeepSeek-V3} technical report.
\newblock \emph{arXiv preprint arXiv:2412.19437}, 2024.
\newblock URL \url{https://arxiv.org/abs/2412.19437}.

\bibitem[DeepSeek-AI(2026)]{deepseek2026v4}
DeepSeek-AI.
\newblock Deepseek-v4: Towards highly efficient million-token context intelligence.
\newblock 2026.
\newblock URL \url{https://huggingface.co/deepseek-ai/DeepSeek-V4-Pro/blob/main/DeepSeek_V4.pdf}.

\bibitem[{DeepSeek-AI} et~al.(2025){DeepSeek-AI}, Guo, Yang, Zhang, Song, Wang, Zhu, Xu, Zhang, Ma, et~al.]{deepseek2025r1}
{DeepSeek-AI}, Daya Guo, Dejian Yang, Haowei Zhang, Junxiao Song, Peiyi Wang, Qihao Zhu, Runxin Xu, Ruoyu Zhang, Shirong Ma, et~al.
\newblock Deepseek-r1 incentivizes reasoning in {LLMs} through reinforcement learning.
\newblock \emph{Nature}, 645:\penalty0 633--638, 2025.
\newblock \doi{10.1038/s41586-025-09422-z}.
\newblock Preprint: arXiv:2501.12948.

\bibitem[Diao et~al.(2025)Diao, Li, Liu, Liao, Wang, Cai, and Xu]{diao2025seas}
Muxi Diao, Rumei Li, Shiyang Liu, Guogang Liao, Jingang Wang, Xunliang Cai, and Weiran Xu.
\newblock {SEAS:} self-evolving adversarial safety optimization for large language models.
\newblock In Toby Walsh, Julie Shah, and Zico Kolter (eds.), \emph{Thirty-Ninth {AAAI} Conference on Artificial Intelligence, Thirty-Seventh Conference on Innovative Applications of Artificial Intelligence, Fifteenth Symposium on Educational Advances in Artificial Intelligence, {AAAI} 2025, Philadelphia, PA, USA, February 25 - March 4, 2025}, pp.\  23778--23786. {AAAI} Press, 2025.
\newblock \doi{10.1609/AAAI.V39I22.34549}.
\newblock URL \url{https://doi.org/10.1609/aaai.v39i22.34549}.

\bibitem[Dragomir et~al.(2026)Dragomir, Brad, and Ionescu]{dragomir2025clewr}
Alexandra Dragomir, Florin Brad, and Radu~Tudor Ionescu.
\newblock Clewr: Curriculum learning with restarts for machine translation preference learning.
\newblock \emph{CoRR}, abs/2601.05858, 2026.
\newblock \doi{10.48550/ARXIV.2601.05858}.
\newblock URL \url{https://doi.org/10.48550/arXiv.2601.05858}.

\bibitem[Dua et~al.(2019)Dua, Wang, Dasigi, Stanovsky, Singh, and Gardner]{dua2019drop}
Dheeru Dua, Yizhong Wang, Pradeep Dasigi, Gabriel Stanovsky, Sameer Singh, and Matt Gardner.
\newblock Drop: A reading comprehension benchmark requiring discrete reasoning over paragraphs.
\newblock \emph{Proceedings of NAACL}, 2019.

\bibitem[ElZemity et~al.(2025)ElZemity, Arief, and Li]{tihanyi2025cyberllminstruct}
Adel ElZemity, Budi Arief, and Shujun Li.
\newblock {CyberLLMInstruct}: A pseudo-malicious dataset revealing safety-performance trade-offs in cyber security {LLM} fine-tuning.
\newblock In \emph{Proceedings of the 18th ACM Workshop on Artificial Intelligence and Security (AISec '25)}, pp.\  77--88, New York, NY, USA, 2025. Association for Computing Machinery.
\newblock ISBN 979-8-4007-1895-3.
\newblock \doi{10.1145/3733799.3762968}.
\newblock Preprint available as arXiv:2503.09334.

\bibitem[{FlagEval}(2023)]{cui2022csem}
{FlagEval}.
\newblock {C-SEM}: Chinese semantic evaluation benchmark dataset.
\newblock FlagEval Benchmark Dataset (v1.0), GitHub Repository, 2023.
\newblock URL \url{https://github.com/flageval-baai/FlagEval/tree/master/csem}.
\newblock Built by Peking University and Minjiang College as co-builders of the FlagEval project, Beijing Academy of Artificial Intelligence (BAAI); four multiple-choice subsets (LLSRC, SLSRC, SLPWC, SLRFC); the FlagEval platform version is larger and evaluated 5-shot.

\bibitem[Gao et~al.(2025)Gao, Wu, Lin, Zhang, and Hu]{gao2025nextlong}
Chaochen Gao, Xing Wu, Zijia Lin, Debing Zhang, and Songlin Hu.
\newblock Nextlong: Toward effective long-context training without long documents.
\newblock In \emph{Proceedings of the Forty-Second International Conference on Machine Learning (ICML)}, 2025.

\bibitem[Ge et~al.(2021)Ge, Sun, Xiong, and Liu]{ge2021wplc}
Huibin Ge, Chenxi Sun, Deyi Xiong, and Qun Liu.
\newblock {C}hinese {WPLC}: A {C}hinese dataset for evaluating pretrained language models on word prediction given long-range context.
\newblock In \emph{Proceedings of the 2021 Conference on Empirical Methods in Natural Language Processing}, pp.\  3770--3778, Online and Punta Cana, Dominican Republic, November 2021. Association for Computational Linguistics.
\newblock \doi{10.18653/v1/2021.emnlp-main.306}.
\newblock URL \url{https://aclanthology.org/2021.emnlp-main.306/}.

\bibitem[Gema et~al.(2025)Gema, Leang, Hong, Devoto, Mancino, Saxena, He, Zhao, Du, Ghasemi~Madani, Barale, McHardy, Harris, Kaddour, Van~Krieken, and Minervini]{garg2024mmluredux}
Aryo~Pradipta Gema, Joshua Ong~Jun Leang, Giwon Hong, Alessio Devoto, Alberto Carlo~Maria Mancino, Rohit Saxena, Xuanli He, Yu~Zhao, Xiaotang Du, Mohammad~Reza Ghasemi~Madani, Claire Barale, Robert McHardy, Joshua Harris, Jean Kaddour, Emile Van~Krieken, and Pasquale Minervini.
\newblock Are we done with {MMLU}?
\newblock In \emph{Proceedings of the 2025 Conference of the Nations of the {A}mericas Chapter of the {A}ssociation for {C}omputational {L}inguistics: Human Language Technologies (Volume 1: Long Papers)}, pp.\  5069--5096, Albuquerque, New Mexico, April 2025. Association for Computational Linguistics.
\newblock ISBN 979-8-89176-189-6.
\newblock \doi{10.18653/v1/2025.naacl-long.262}.
\newblock URL \url{https://aclanthology.org/2025.naacl-long.262/}.

\bibitem[Grattafiori et~al.(2024)]{grattafiori2024llama3}
Aaron Grattafiori et~al.
\newblock The {Llama} 3 herd of models.
\newblock \emph{arXiv preprint arXiv:2407.21783}, 2024.
\newblock URL \url{https://arxiv.org/abs/2407.21783}.

\bibitem[Gu et~al.(2024)Gu, Zhu, Ye, Zhang, Wang, Zhu, Jiang, Xiong, Li, Wu, He, Xu, Huang, Liu, Wang, Wang, Zheng, Feng, and Xiao]{gu2023xiezhi}
Zhouhong Gu, Xiaoxuan Zhu, Haoning Ye, Lin Zhang, Jianchen Wang, Yixin Zhu, Sihang Jiang, Zhuozhi Xiong, Zihan Li, Weijie Wu, Qianyu He, Rui Xu, Wenhao Huang, Jingping Liu, Zili Wang, Shusen Wang, Weiguo Zheng, Hongwei Feng, and Yanghua Xiao.
\newblock {Xiezhi}: An ever-updating benchmark for holistic domain knowledge evaluation.
\newblock In \emph{Proceedings of the AAAI Conference on Artificial Intelligence}, volume~38, pp.\  18099--18107, 2024.
\newblock \doi{10.1609/aaai.v38i16.29767}.
\newblock arXiv:2306.05783.

\bibitem[Guo et~al.(2025)]{guo2025deepseekr1}
Daya Guo et~al.
\newblock Deepseek-r1 incentivizes reasoning in {LLMs} through reinforcement learning.
\newblock \emph{Nature}, 645:\penalty0 633--638, 2025.
\newblock \doi{10.1038/s41586-025-09422-z}.
\newblock URL \url{https://doi.org/10.1038/s41586-025-09422-z}.
\newblock Preprint: arXiv:2501.12948.

\bibitem[Hendrycks et~al.(2021{\natexlab{a}})Hendrycks, Burns, Basart, Zou, Mazeika, Song, and Steinhardt]{hendrycks2021mmlu}
Dan Hendrycks, Collin Burns, Steven Basart, Andy Zou, Mantas Mazeika, Dawn Song, and Jacob Steinhardt.
\newblock Measuring massive multitask language understanding.
\newblock In \emph{International Conference on Learning Representations}, 2021{\natexlab{a}}.

\bibitem[Hendrycks et~al.(2021{\natexlab{b}})Hendrycks, Burns, Kadavath, Arora, Basart, Tang, Song, and Steinhardt]{hendrycks2021math}
Dan Hendrycks, Collin Burns, Saurav Kadavath, Akul Arora, Steven Basart, Eric Tang, Dawn Song, and Jacob Steinhardt.
\newblock Measuring mathematical problem solving with the math dataset.
\newblock \emph{Proceedings of NeurIPS}, 2021{\natexlab{b}}.

\bibitem[Hu et~al.(2020)Hu, Richardson, Xu, Li, K{\"u}bler, and Moss]{hu2020ocnli}
Hai Hu, Kyle Richardson, Liang Xu, Lu~Li, Sandra K{\"u}bler, and Lawrence Moss.
\newblock {OCNLI}: {O}riginal {C}hinese {N}atural {L}anguage {I}nference.
\newblock In \emph{Findings of the {A}ssociation for {C}omputational {L}inguistics: {EMNLP} 2020}, pp.\  3512--3526, Online, November 2020. Association for Computational Linguistics.
\newblock \doi{10.18653/v1/2020.findings-emnlp.314}.
\newblock URL \url{https://aclanthology.org/2020.findings-emnlp.314/}.

\bibitem[Huang et~al.(2023)Huang, Bai, Zhu, Zhang, Zhang, Su, Liu, Lv, Zhang, Lei, Fu, Sun, and He]{huang2023ceval}
Yuzhen Huang, Yuzhuo Bai, Zhihao Zhu, Junlei Zhang, Jinghan Zhang, Tangjun Su, Junteng Liu, Chuancheng Lv, Yikai Zhang, Jiayi Lei, Yao Fu, Maosong Sun, and Junxian He.
\newblock {C-Eval}: A multi-level multi-discipline chinese evaluation suite for foundation models.
\newblock In \emph{Advances in Neural Information Processing Systems 36 (NeurIPS 2023) Datasets and Benchmarks Track}, volume~36, pp.\  62991--63010. Curran Associates, Inc., 2023.
\newblock URL \url{https://proceedings.neurips.cc/paper_files/paper/2023/hash/c6ec1844bec96d6d32ae95ae694e23d8-Abstract-Datasets_and_Benchmarks.html}.

\bibitem[Jaech et~al.(2024)]{jaech2024o1}
Aaron Jaech et~al.
\newblock {OpenAI o1} system card.
\newblock \emph{arXiv preprint arXiv:2412.16720}, 2024.
\newblock URL \url{https://arxiv.org/abs/2412.16720}.

\bibitem[Jia et~al.(2025)Jia, Chen, Wu, Gao, Lin, Zhang, Hu, and Guo]{jia2025entropylong}
Junlong Jia, Ziyang Chen, Xing Wu, Chaochen Gao, Zijia Lin, Debing Zhang, Songlin Hu, and Binghui Guo.
\newblock Entropylong: Effective long-context training via predictive uncertainty.
\newblock \emph{CoRR}, abs/2510.02330, 2025.
\newblock \doi{10.48550/ARXIV.2510.02330}.
\newblock URL \url{https://doi.org/10.48550/arXiv.2510.02330}.

\bibitem[Jing et~al.(2024)Jing, Tang, Shi, Zheng, Nie, Wu, Yang, and Luo]{liu2024secbench}
Pengfei Jing, Mengyun Tang, Xiaorong Shi, Xing Zheng, Sen Nie, Shi Wu, Yong Yang, and Xiapu Luo.
\newblock {SecBench}: A comprehensive multi-dimensional benchmarking dataset for {LLMs} in cybersecurity.
\newblock \emph{arXiv preprint arXiv:2412.20787}, 2024.
\newblock \doi{10.48550/arXiv.2412.20787}.
\newblock URL \url{https://arxiv.org/abs/2412.20787}.

\bibitem[Joshi et~al.(2017)Joshi, Choi, Weld, and Zettlemoyer]{joshi2017triviaqa}
Mandar Joshi, Eunsol Choi, Daniel~S Weld, and Luke Zettlemoyer.
\newblock Triviaqa: A large scale distantly supervised challenge dataset for reading comprehension.
\newblock \emph{Proceedings of ACL}, 2017.

\bibitem[Kassianik et~al.(2025)]{kassianik2025foundationsec8b}
Paul Kassianik et~al.
\newblock {Llama-3.1-FoundationAI-SecurityLLM-Base-8B} technical report.
\newblock \emph{arXiv preprint arXiv:2504.21039}, 2025.
\newblock URL \url{https://arxiv.org/abs/2504.21039}.

\bibitem[Kiraz(2025)]{baronsecurity2025baronllm}
Alican Kiraz.
\newblock {BaronLLM}: Offensive security language model.
\newblock HuggingFace Model Repository, 2025.
\newblock URL \url{https://huggingface.co/AlicanKiraz0/Cybersecurity-BaronLLM_Offensive_Security_LLM_Q6_K_GGUF}.
\newblock Llama-3.1-8B-Instruct fine-tune for offensive security research, published 21 January 2025; the later BaronLLM v2 (Qwen3-14B, August 2025) is a separate model from the Trendyol Group Security Team.

\bibitem[Ko{\v{c}}isk{\'y} et~al.(2018)Ko{\v{c}}isk{\'y}, Schwarz, Blunsom, Dyer, Hermann, Melis, and Grefenstette]{kocisky2018narrativeqa}
Tom{\'a}{\v{s}} Ko{\v{c}}isk{\'y}, Jonathan Schwarz, Phil Blunsom, Chris Dyer, Karl~Moritz Hermann, G{\'a}bor Melis, and Edward Grefenstette.
\newblock The narrativeqa reading comprehension challenge.
\newblock \emph{Transactions of the ACL}, 2018.

\bibitem[Li et~al.(2023)Li, Li, Wang, Yang, and Yu]{li2023seceval}
Guancheng Li, Yifeng Li, Guannan Wang, Haoyu Yang, and Yang Yu.
\newblock {SecEval}: A comprehensive benchmark for evaluating cybersecurity knowledge of foundation models, 2023.
\newblock URL \url{https://github.com/XuanwuAI/SecEval}.

\bibitem[Li et~al.(2024)Li, Zhang, Koto, Yang, Zhao, Gong, Duan, and Baldwin]{li2023cmmlu}
Haonan Li, Yixuan Zhang, Fajri Koto, Yifei Yang, Hai Zhao, Yeyun Gong, Nan Duan, and Timothy Baldwin.
\newblock {CMMLU}: Measuring massive multitask language understanding in {C}hinese.
\newblock In \emph{Findings of the {A}ssociation for {C}omputational {L}inguistics: {ACL} 2024}, pp.\  11260--11285, Bangkok, Thailand, August 2024. Association for Computational Linguistics.
\newblock \doi{10.18653/v1/2024.findings-acl.671}.
\newblock URL \url{https://aclanthology.org/2024.findings-acl.671/}.

\bibitem[Liu et~al.(2024)Liu, Zeng, He, Jiang, and He]{liu2024deita}
Wei Liu, Weihao Zeng, Keqing He, Yong Jiang, and Junxian He.
\newblock What makes good data for alignment? a comprehensive study of automatic data selection in instruction tuning.
\newblock In \emph{Proceedings of the Twelfth International Conference on Learning Representations (ICLR)}, 2024.

\bibitem[Lu et~al.(2026)Lu, He, Chen, and Zha]{mssr2026}
Yiyang Lu, Yu~He, Jianlong Chen, and Hongyuan Zha.
\newblock {MSSR}: Memory-aware adaptive replay for continual {LLM} fine-tuning.
\newblock \emph{arXiv preprint arXiv:2603.09892}, 2026.

\bibitem[Mihaylov et~al.(2018)Mihaylov, Clark, Khot, and Sabharwal]{mihaylov2018openbookqa}
Todor Mihaylov, Peter Clark, Tushar Khot, and Ashish Sabharwal.
\newblock Can a suit of armor conduct electricity? a new dataset for open book question answering.
\newblock \emph{Proceedings of EMNLP}, 2018.

\bibitem[Mostafazadeh et~al.(2016)Mostafazadeh, Chambers, He, Parikh, Batra, Vanderwende, Kohli, and Allen]{mostafazadeh2017storycloze}
Nasrin Mostafazadeh, Nathanael Chambers, Xiaodong He, Devi Parikh, Dhruv Batra, Lucy Vanderwende, Pushmeet Kohli, and James Allen.
\newblock A corpus and cloze evaluation for deeper understanding of commonsense stories.
\newblock In Kevin Knight, Ani Nenkova, and Owen Rambow (eds.), \emph{Proceedings of the 2016 Conference of the {N}orth {A}merican Chapter of the {A}ssociation for {C}omputational {L}inguistics: Human Language Technologies}, pp.\  839--849, San Diego, California, June 2016. Association for Computational Linguistics.
\newblock \doi{10.18653/v1/N16-1098}.
\newblock URL \url{https://aclanthology.org/N16-1098/}.

\bibitem[{OpenAI}(2026)]{openai2026gpt54cyber}
{OpenAI}.
\newblock Scaling trusted access for the next era of cyber defense, 2026.
\newblock URL \url{https://openai.com/index/scaling-trusted-access-for-cyber-defense/}.

\bibitem[Paperno et~al.(2016)Paperno, Kruszewski, Lazaridou, Pham, Bernardi, Pezzelle, Baroni, Boleda, and Fern{\'a}ndez]{paperno2016lambada}
Denis Paperno, Germ{\'a}n Kruszewski, Angeliki Lazaridou, Ngoc~Quan Pham, Raffaella Bernardi, Sandro Pezzelle, Marco Baroni, Gemma Boleda, and Raquel Fern{\'a}ndez.
\newblock The {LAMBADA} dataset: Word prediction requiring a broad discourse context.
\newblock In \emph{Proceedings of the 54th Annual Meeting of the {A}ssociation for {C}omputational {L}inguistics (Volume 1: Long Papers)}, pp.\  1525--1534, Berlin, Germany, August 2016. Association for Computational Linguistics.
\newblock \doi{10.18653/v1/2016.acl-long.1144}.
\newblock URL \url{https://aclanthology.org/P16-1144/}.

\bibitem[{Qoder Team}(2026)]{qoder2026actionrl}
{Qoder Team}.
\newblock The next evolution toward intelligent editing: {Qoder} {NEXT} model and {ActionRL} preference alignment in practice, 2026.
\newblock URL \url{https://qoder.com/blog/qoder-next-model?type=%E6%8A%80%E6%9C%AF}.
\newblock Accessed: 2025.

\bibitem[{Qwen Team}(2025)]{qwenteam2025qwen3}
{Qwen Team}.
\newblock Qwen3 technical report.
\newblock \emph{arXiv preprint arXiv:2505.09388}, 2025.
\newblock URL \url{https://arxiv.org/abs/2505.09388}.

\bibitem[{Qwen Team}(2026)]{qwenteam2025qwen35}
{Qwen Team}.
\newblock {Qwen3.5}: Towards native multimodal agents.
\newblock Qwen Team Blog Post, February 2026.
\newblock URL \url{https://qwen.ai/blog?id=qwen3.5}.
\newblock Official release announcement for the Qwen3.5 open model series (0.8B to 397B-A17B); no standalone Qwen3.5 technical report has been published. Alibaba Cloud mirror: \url{https://www.alibabacloud.com/blog/602894}.

\bibitem[Rafailov et~al.(2023)Rafailov, Sharma, Mitchell, Manning, Ermon, and Finn]{dpo23}
Rafael Rafailov, Archit Sharma, Eric Mitchell, Christopher~D. Manning, Stefano Ermon, and Chelsea Finn.
\newblock Direct preference optimization: Your language model is secretly a reward model.
\newblock In Alice Oh, Tristan Naumann, Amir Globerson, Kate Saenko, Moritz Hardt, and Sergey Levine (eds.), \emph{Advances in Neural Information Processing Systems 36: Annual Conference on Neural Information Processing Systems 2023, NeurIPS 2023, New Orleans, LA, USA, December 10 - 16, 2023}, 2023.
\newblock URL \url{http://papers.nips.cc/paper\_files/paper/2023/hash/a85b405ed65c6477a4fe8302b5e06ce7-Abstract-Conference.html}.

\bibitem[Rajpurkar et~al.(2018)Rajpurkar, Jia, and Liang]{rajpurkar2018squad}
Pranav Rajpurkar, Robin Jia, and Percy Liang.
\newblock Know what you don't know: Unanswerable questions for squad.
\newblock \emph{Proceedings of ACL}, 2018.

\bibitem[Rein et~al.(2024)Rein, Hou, Stickland, Petty, Pang, Dirani, Michael, and Bowman]{rein2024gpqa}
David Rein, Betty Hou, Asa Stickland, Jackson Petty, Richard~Y Pang, Julien Dirani, Julian Michael, and Samuel~R Bowman.
\newblock Gpqa: A graduate-level google-proof q\&a benchmark.
\newblock \emph{Proceedings of ICLR}, 2024.

\bibitem[Sakaguchi et~al.(2020)Sakaguchi, Le~Bras, Bhagavatula, and Choi]{sakaguchi2021winogrande}
Keisuke Sakaguchi, Ronan Le~Bras, Chandra Bhagavatula, and Yejin Choi.
\newblock {WinoGrande}: An adversarial winograd schema challenge at scale.
\newblock In \emph{Proceedings of the AAAI Conference on Artificial Intelligence}, volume~34, pp.\  8732--8740. AAAI Press, 2020.
\newblock \doi{10.1609/aaai.v34i05.6399}.
\newblock URL \url{https://doi.org/10.1609/aaai.v34i05.6399}.

\bibitem[Sap et~al.(2019)Sap, Rashkin, Chen, Le~Bras, and Choi]{sap2019socialiqa}
Maarten Sap, Hannah Rashkin, Derek Chen, Ronan Le~Bras, and Yejin Choi.
\newblock Social iqa: Commonsense reasoning about social interactions.
\newblock \emph{Proceedings of EMNLP}, 2019.

\bibitem[Schaul et~al.(2016)Schaul, Quan, Antonoglou, and Silver]{schaul2015per}
Tom Schaul, John Quan, Ioannis Antonoglou, and David Silver.
\newblock Prioritized experience replay.
\newblock In Yoshua Bengio and Yann LeCun (eds.), \emph{4th International Conference on Learning Representations, {ICLR} 2016, San Juan, Puerto Rico, May 2-4, 2016, Conference Track Proceedings}, 2016.
\newblock URL \url{http://arxiv.org/abs/1511.05952}.

\bibitem[{segolilylabs}(2025)]{segolilylabs2025lily}
{segolilylabs}.
\newblock Lily-cybersecurity-7b-v0.2.
\newblock HuggingFace Model Repository, 2025.
\newblock URL \url{https://huggingface.co/segolilylabs/Lily-Cybersecurity-7B-v0.2}.

\bibitem[Shao et~al.(2024)]{shao2024grpo}
Zhihong Shao et~al.
\newblock {DeepSeekMath}: Pushing the limits of mathematical reasoning in open language models.
\newblock \emph{arXiv preprint arXiv:2402.03300}, 2024.
\newblock URL \url{https://arxiv.org/abs/2402.03300}.
\newblock Introduces Group Relative Policy Optimization (GRPO).

\bibitem[Sun et~al.(2020)Sun, Yu, Yu, and Cardie]{sun2020c3}
Kai Sun, Dian Yu, Dong Yu, and Claire Cardie.
\newblock Investigating prior knowledge for challenging {C}hinese machine reading comprehension.
\newblock \emph{Transactions of the Association for Computational Linguistics}, 8:\penalty0 141--155, 2020.
\newblock \doi{10.1162/tacl_a_00305}.
\newblock URL \url{https://aclanthology.org/2020.tacl-1.10/}.

\bibitem[Suzgun et~al.(2023)Suzgun, Scales, Sch{\"a}rli, Gehrmann, Tay, Chung, Chowdhery, Le, Chi, Zhou, and Wei]{suzgun2023bbh}
Mirac Suzgun, Nathan Scales, Nathanael Sch{\"a}rli, Sebastian Gehrmann, Yi~Tay, Hyung~Won Chung, Aakanksha Chowdhery, Quoc~V. Le, Ed~H. Chi, Denny Zhou, and Jason Wei.
\newblock Challenging {BIG}-bench tasks and whether chain-of-thought can solve them.
\newblock In \emph{Findings of the {A}ssociation for {C}omputational {L}inguistics: {ACL} 2023}, pp.\  13003--13051, Toronto, Canada, July 2023. Association for Computational Linguistics.
\newblock \doi{10.18653/v1/2023.findings-acl.824}.
\newblock URL \url{https://aclanthology.org/2023.findings-acl.824/}.

\bibitem[{TAL Education Group}(2023)]{zheng2021tal}
{TAL Education Group}.
\newblock {TAL-SCQ5K}: A high-quality mathematical competition dataset in english and chinese.
\newblock HuggingFace Dataset Repository, 2023.
\newblock URL \url{https://huggingface.co/datasets/math-eval/TAL-SCQ5K}.
\newblock {TAL-SCQ5K-EN} and {TAL-SCQ5K-CN} splits of 5K single-choice questions each (3K train / 2K test); MIT License; no accompanying paper. Mirror repository: \url{https://github.com/math-eval/TAL-SCQ5K}.

\bibitem[Talmor et~al.(2019)Talmor, Herzig, Lourie, and Berant]{talmor2019commonsenseqa}
Alon Talmor, Jonathan Herzig, Nicholas Lourie, and Jonathan Berant.
\newblock Commonsenseqa: A question answering challenge targeting commonsense knowledge.
\newblock \emph{Proceedings of NAACL}, 2019.

\bibitem[Tihanyi et~al.(2024)Tihanyi, Ferrag, Jain, Bisztray, and Debbah]{tihanyi2024cybermetric}
Norbert Tihanyi, Mohamed~Amine Ferrag, Ridhi Jain, Tamas Bisztray, and Merouane Debbah.
\newblock {CyberMetric}: A benchmark dataset based on retrieval-augmented generation for evaluating {LLMs} in cybersecurity knowledge.
\newblock In \emph{2024 IEEE International Conference on Cyber Security and Resilience (CSR)}, pp.\  296--302, 2024.

\bibitem[Wang et~al.(2026)]{wang2026codea1}
Aozhe Wang et~al.
\newblock {Code-A1}: Adversarial evolving of code {LLM} and test {LLM} via reinforcement learning.
\newblock \emph{arXiv preprint arXiv:2603.15611}, 2026.

\bibitem[Wang et~al.(2025{\natexlab{a}})Wang, Zhou, Li, Bai, Chen, Qin, Sun, and Li]{wang2025csebenchmark}
Dawei Wang, Geng Zhou, Xianglong Li, Yu~Bai, Li~Chen, Ting Qin, Jian Sun, and Dan Li.
\newblock The digital cybersecurity expert: How far have we come?
\newblock In \emph{2025 IEEE Symposium on Security and Privacy (SP)}, pp.\  3273--3290, 2025{\natexlab{a}}.
\newblock \doi{10.1109/SP61157.2025.00198}.

\bibitem[Wang et~al.(2025{\natexlab{b}})Wang, Yang, Wang, Wang, Xu, and Huang]{wang2025abkd}
Guanghui Wang, Zhiyong Yang, Zitai Wang, Shi Wang, Qianqian Xu, and Qingming Huang.
\newblock {ABKD}: Pursuing a proper allocation of the probability mass in knowledge distillation via $\alpha$-$\beta$-divergence.
\newblock In \emph{Proceedings of the 42nd International Conference on Machine Learning}, volume 267 of \emph{Proceedings of Machine Learning Research}, pp.\  65167--65212. PMLR, 2025{\natexlab{b}}.

\bibitem[Wang et~al.(2023)Wang, Kordi, Mishra, Liu, Smith, Khashabi, and Hajishirzi]{wang2023selfinstruct}
Yizhong Wang, Yeganeh Kordi, Swaroop Mishra, Alisa Liu, Noah~A Smith, Daniel Khashabi, and Hannaneh Hajishirzi.
\newblock Self-instruct: Aligning language models with self-generated instructions.
\newblock In \emph{Proceedings of the 61st Annual Meeting of the Association for Computational Linguistics (ACL)}, 2023.

\bibitem[Wang et~al.(2024)Wang, Ma, Zhang, Ni, Chandra, Guo, Ren, Arulraj, He, Jiang, Li, Ku, Wang, Zhuang, Fan, Yue, and Chen]{wang2024mmlupro}
Yubo Wang, Xueguang Ma, Ge~Zhang, Yuansheng Ni, Abhranil Chandra, Shiguang Guo, Weiming Ren, Aaran Arulraj, Xuan He, Ziyan Jiang, Tianle Li, Max Ku, Kai Wang, Alex Zhuang, Rongqi Fan, Xiang Yue, and Wenhu Chen.
\newblock {MMLU-Pro}: A more robust and challenging multi-task language understanding benchmark.
\newblock In \emph{Advances in Neural Information Processing Systems 37 (NeurIPS 2024) Datasets and Benchmarks Track}, volume~37, pp.\  95266--95290. Curran Associates, Inc., 2024.
\newblock URL \url{https://proceedings.neurips.cc/paper_files/paper/2024/hash/ad236edc564f3e3156e1b2feafb99a24-Abstract-Datasets_and_Benchmarks_Track.html}.

\bibitem[Wei et~al.(2025)Wei, Duchenne, Copet, Carbonneaux, Zhang, Fried, Synnaeve, Singh, and Wang]{wei2025swerl}
Yuxiang Wei, Olivier Duchenne, Jade Copet, Quentin Carbonneaux, Lingming Zhang, Daniel Fried, Gabriel Synnaeve, Rishabh Singh, and Sida~I. Wang.
\newblock {SWE-RL}: Advancing llm reasoning via reinforcement learning on open software evolution.
\newblock In \emph{Advances in Neural Information Processing Systems (NeurIPS)}, 2025.

\bibitem[Wortsman et~al.(2022)Wortsman, Ilharco, Gadre, Roelofs, Gontijo-Lopes, Morcos, Namkoong, Farhadi, Carmon, Kornblith, and Schmidt]{wortsman2022model}
Mitchell Wortsman, Gabriel Ilharco, Samir~Yitzhak Gadre, Rebecca Roelofs, Raphael Gontijo-Lopes, Ari~S. Morcos, Hongseok Namkoong, Ali Farhadi, Yair Carmon, Simon Kornblith, and Ludwig Schmidt.
\newblock Model soups: Averaging weights of multiple fine-tuned models improves accuracy without increasing inference time.
\newblock In \emph{International Conference on Machine Learning (ICML)}, pp.\  23965--23998, 2022.

\bibitem[Xie et~al.(2023)Xie, Pham, Dong, Du, Liu, Lu, Liang, Le, Ma, and Yu]{xie2023doremi}
Sang~Michael Xie, Hieu Pham, Xuanyi Dong, Nan Du, Hanxiao Liu, Yifeng Lu, Percy Liang, Quoc~V. Le, Tengyu Ma, and Adams~Wei Yu.
\newblock Doremi: Optimizing data mixtures speeds up language model pretraining.
\newblock In \emph{Advances in Neural Information Processing Systems}, 2023.

\bibitem[Xu et~al.(2025)Xu, Wang, Li, Wang, Zhao, Chen, Yu, Liu, and Wang]{xu2024llmcybersecurity}
Hanxiang Xu, Shenao Wang, Ningke Li, Kailong Wang, Yanjie Zhao, Kai Chen, Ting Yu, Yang Liu, and Haoyu Wang.
\newblock Large language models for cyber security: A systematic literature review.
\newblock \emph{ACM Trans. Softw. Eng. Methodol.}, September 2025.
\newblock ISSN 1049-331X.
\newblock \doi{10.1145/3769676}.
\newblock URL \url{https://doi.org/10.1145/3769676}.

\bibitem[Xu et~al.(2020)Xu, Hu, Zhang, Li, Cao, Li, Xu, Sun, Yu, Yu, Tian, Dong, Liu, Shi, Cui, Li, Zeng, Wang, Xie, Li, Patterson, Tian, Zhang, Zhou, Liu, Zhao, Zhao, Yue, Zhang, Yang, Richardson, and Lan]{xu2020clue}
Liang Xu, Hai Hu, Xuanwei Zhang, Lu~Li, Chenjie Cao, Yudong Li, Yechen Xu, Kai Sun, Dian Yu, Cong Yu, Yin Tian, Qianqian Dong, Weitang Liu, Bo~Shi, Yiming Cui, Junyi Li, Jun Zeng, Rongzhao Wang, Weijian Xie, Yanting Li, Yina Patterson, Zuoyu Tian, Yiwen Zhang, He~Zhou, Shaoweihua Liu, Zhe Zhao, Qipeng Zhao, Cong Yue, Xinrui Zhang, Zhengliang Yang, Kyle Richardson, and Zhenzhong Lan.
\newblock {CLUE}: A {C}hinese language understanding evaluation benchmark.
\newblock In Donia Scott, Nuria Bel, and Chengqing Zong (eds.), \emph{Proceedings of the 28th International Conference on Computational Linguistics}, pp.\  4762--4772, Barcelona, Spain (Online), December 2020. International Committee on Computational Linguistics.
\newblock \doi{10.18653/v1/2020.coling-main.419}.
\newblock URL \url{https://aclanthology.org/2020.coling-main.419/}.

\bibitem[Yadav et~al.(2023)Yadav, Tam, Choshen, Raffel, and Bansal]{yadav2024ties}
Prateek Yadav, Derek Tam, Leshem Choshen, Colin Raffel, and Mohit Bansal.
\newblock {TIES}-merging: Resolving interference when merging models.
\newblock In \emph{Advances in Neural Information Processing Systems (NeurIPS)}, 2023.

\bibitem[Yang et~al.(2024)]{yang2024qwen2}
An~Yang et~al.
\newblock Qwen2 technical report.
\newblock \emph{arXiv preprint arXiv:2407.10671}, 2024.
\newblock URL \url{https://arxiv.org/abs/2407.10671}.

\bibitem[Yang et~al.(2026)Yang, Li, He, Priyanshu, Saglam, Kassianik, Weerawardhena, Vellore, Nelson, Javidnia, Goldblatt, Burch, Zohary, Eisenman, Sabbaghi, Vijay, Dharssi, Kedia, Oshiba, Singer, and Karbasi]{yang2026foundationsec8breasoning}
Zhuoran Yang, Ed~Li, Jianliang He, Aman Priyanshu, Baturay Saglam, Paul Kassianik, Sajana Weerawardhena, Anu Vellore, Blaine Nelson, Neusha Javidnia, Arthur Goldblatt, Fraser Burch, Avi Zohary, Assaf Eisenman, Mahdi Sabbaghi, Supriti Vijay, Rahim Dharssi, Dhruv Kedia, Kojin Oshiba, Yaron Singer, and Amin Karbasi.
\newblock {Llama-3.1-FoundationAI-SecurityLLM-Reasoning-8B Technical Report}, 2026.
\newblock URL \url{https://arxiv.org/abs/2601.21051}.

\bibitem[Yen et~al.(2024)Yen, Gao, Hou, Ding, Fleischer, Izsak, Wasserblat, and Chen]{yen2024helmet}
Howard Yen, Tianyu Gao, Minmin Hou, Ke~Ding, Daniel Fleischer, Peter Izsak, Moshe Wasserblat, and Danqi Chen.
\newblock Helmet: How to evaluate long-context language models effectively and thoroughly.
\newblock \emph{arXiv preprint arXiv:2410.02694}, 2024.

\bibitem[Yu et~al.(2020)Yu, Kumar, Gupta, Levine, Hausman, and Finn]{yu2020pcgrad}
Tianhe Yu, Saurabh Kumar, Abhishek Gupta, Sergey Levine, Karol Hausman, and Chelsea Finn.
\newblock Gradient surgery for multi-task learning.
\newblock In Hugo Larochelle, Marc'Aurelio Ranzato, Raia Hadsell, Maria{-}Florina Balcan, and Hsuan{-}Tien Lin (eds.), \emph{Advances in Neural Information Processing Systems 33: Annual Conference on Neural Information Processing Systems 2020, NeurIPS 2020, December 6-12, 2020, virtual}, 2020.
\newblock URL \url{https://proceedings.neurips.cc/paper/2020/hash/3fe78a8acf5fda99de95303940a2420c-Abstract.html}.

\bibitem[Yu et~al.(2025)Yu, Chiang, Tsai, Huang, and Tsao]{yu2025primus}
Yao{-}Ching Yu, Tsun{-}Han Chiang, Cheng{-}Wei Tsai, Chien{-}Ming Huang, and Wen{-}Kwang Tsao.
\newblock Primus: {A} pioneering collection of open-source datasets for cybersecurity {LLM} training.
\newblock In Christos Christodoulopoulos, Tanmoy Chakraborty, Carolyn Rose, and Violet Peng (eds.), \emph{Proceedings of the 2025 Conference on Empirical Methods in Natural Language Processing, {EMNLP} 2025, Suzhou, China, November 4-9, 2025}, pp.\  10391--10413. Association for Computational Linguistics, 2025.
\newblock \doi{10.18653/V1/2025.EMNLP-MAIN.527}.
\newblock URL \url{https://doi.org/10.18653/v1/2025.emnlp-main.527}.

\bibitem[Zellers et~al.(2019)Zellers, Holtzman, Bisk, Farhadi, and Choi]{zellers2019hellaswag}
Rowan Zellers, Ari Holtzman, Yonatan Bisk, Ali Farhadi, and Yejin Choi.
\newblock Hellaswag: Can a machine really finish your sentence?
\newblock \emph{Proceedings of ACL}, 2019.

\bibitem[Zeng et~al.(2026)Zeng, Lv, Hou, Du, Zheng, Chen, Yin, Ge, Huang, Xie, et~al.]{thudm2026glm5}
Aohan Zeng, Xin Lv, Zhenyu Hou, Zhengxiao Du, Qinkai Zheng, Bin Chen, Da~Yin, Chendi Ge, Chenghua Huang, Chengxing Xie, et~al.
\newblock Glm-5: from vibe coding to agentic engineering.
\newblock \emph{arXiv preprint arXiv:2602.15763}, 2026.
\newblock URL \url{https://arxiv.org/abs/2602.15763}.

\bibitem[Zhang et~al.(2023)Zhang, Wen, Deng, Xin, Li, Li, Zhu, and Sun]{jiang2024hackmentor}
Jie Zhang, Hui Wen, Liting Deng, Mingfeng Xin, Zhi Li, Lun Li, Hongsong Zhu, and Limin Sun.
\newblock {HackMentor}: Fine-tuning large language models for cybersecurity.
\newblock In \emph{Proceedings of the 2023 IEEE 22nd International Conference on Trust, Security and Privacy in Computing and Communications (TrustCom)}, pp.\  452--461, Exeter, United Kingdom, 2023. IEEE.
\newblock \doi{10.1109/TrustCom60117.2023.00076}.

\bibitem[Zhang et~al.(2025)Zhang, Bu, Wen, Liu, Fei, Xi, Li, Yang, Zhu, and Meng]{zhang2024llmcybersecurity}
Jie Zhang, Haoyu Bu, Hui Wen, Yongji Liu, Haiqiang Fei, Rongrong Xi, Lun Li, Yun Yang, Hongsong Zhu, and Dan Meng.
\newblock When llms meet cybersecurity: a systematic literature review.
\newblock \emph{Cybersecur.}, 8\penalty0 (1):\penalty0 55, 2025.
\newblock \doi{10.1186/S42400-025-00361-W}.
\newblock URL \url{https://doi.org/10.1186/s42400-025-00361-w}.

\bibitem[Zhao et~al.(2020)Zhao, Shang, Liu, Wang, and Liu]{zhao2020ape210k}
Wei Zhao, Mingyue Shang, Yang Liu, Liang Wang, and Jingming Liu.
\newblock {Ape210K}: A large-scale and template-rich dataset of math word problems, 2020.
\newblock URL \url{https://arxiv.org/abs/2009.11506}.
\newblock Withdrawn by the authors in v2 (9 October 2020); dataset available at \url{https://github.com/yuantiku/ape210k}.

\bibitem[Zheng et~al.(2019)Zheng, Huang, and Sun]{zheng2020chid}
Chujie Zheng, Minlie Huang, and Aixin Sun.
\newblock {C}h{ID}: A large-scale {C}hinese {ID}iom dataset for cloze test.
\newblock In Anna Korhonen, David Traum, and Llu{\'i}s M{\`a}rquez (eds.), \emph{Proceedings of the 57th Annual Meeting of the {A}ssociation for {C}omputational {L}inguistics}, pp.\  778--787, Florence, Italy, July 2019. Association for Computational Linguistics.
\newblock \doi{10.18653/v1/P19-1075}.
\newblock URL \url{https://aclanthology.org/P19-1075/}.

\bibitem[Zhuo et~al.(2025)Zhuo, Wang, Ding, Kumar, and Wang]{zhuo2025cyberzero}
Terry~Yue Zhuo, Dingmin Wang, Hantian Ding, Varun Kumar, and Zijian Wang.
\newblock Cyber-zero: Training cybersecurity agents without runtime.
\newblock \emph{arXiv preprint arXiv:2508.00910}, 2025.

\bibitem[{ZySec-AI }(2024)]{pentera2025securityllm}
{ZySec-AI }.
\newblock Zysec-7b.
\newblock HuggingFace Model Repository, 2024.
\newblock URL \url{https://huggingface.co/ZySec-AI/SecurityLLM}.

\end{thebibliography}
\bibliographystyle{colm2024_conference}

\clearpage

\end{CJK}
\end{document}